\documentclass[useAMS,usenatbib]{mnras}
\usepackage{graphicx}
\usepackage{amsmath,amssymb}
\newcommand{\f}{\frac}

\newcommand{\BF}{\begin{figure}\begin{center}}
\newcommand{\EF}{\end{center}\end{figure}}
\newcommand{\BE}{\begin{equation}}
\newcommand{\EE}{\end{equation}}
\newcommand{\BEA}{\begin{eqnarray}}
\newcommand{\EEA}{\end{eqnarray}}
\newcommand{\ti}{\textit}
\newcommand{\tr}{\textrm}
\newcommand{\IG}{\includegraphics}
\def\v#1{\boldsymbol #1}
\newcommand{\bvec}[1]{\mbox{\boldmath $#1$}}
\newcommand{\ms}{M_{\odot}}

\begin{document}
\title{ALMA Imprint of Intergalactic Dark Structures in the
Gravitational Lens SDP.81}
\author[Kaiki Taro Inoue, Takeo Minezaki, Satoki Matsushita, Masashi Chiba]
{Kaiki Taro Inoue$^1$\thanks{E-mail:kinoue@phys.kindai.ac.jp}, Takeo
Minezaki$^2$, Satoki Matsushita$^3$, Masashi Chiba$^4$
\\
$^{1}$Faculty of Science and Engineering, 
Kinki University, Higashi-Osaka, 577-8502, Japan 
\\
$^{2}$Institute of Astronomy, School of Science, University of
Tokyo, Mitaka, Tokyo 181-0015, Japan
\\
$^{3}$Academia Sinica Institute of Astronomy and Astrophysics, 
P.O. Box 23-141, Taipei 10617, Taiwan, Republic of China
\\
$^{4}$Astronomical Institute, Tohoku University,
Aoba-ku, Sendai 980-8578, Japan}

%\author{Kaiki Taro Inoue\altaffilmark{1}, Ryuichi Takahashi \altaffilmark{2}}
%\altaffiltext{1}{Department of Science and Engineering, 
%Kinki University, Higashi-Osaka, Osaka 577-8502, Japan}
%\altaffiltext{2}{Faculty of Science and Technology, Hirosaki University, Hirosa%ki, Aomori 036-8561, Japan}

%%
%%
%%
%%
%%
\date{\today}

\pagerange{\pageref{firstpage}--\pageref{lastpage}} \pubyear{0000}

\maketitle

\label{firstpage}
\begin{abstract}
We present an analysis of the ALMA long baseline science verification
 data of the gravitational lens system SDP.81. We fit the positions of 
 the brightest clumps at redshift $z=3.042$ and a possible active
 galactic nucleus component of
 the lensing galaxy at redshift $z=0.2999$ in 
the band 7 continuum image using a canonical lens model,
a singular isothermal ellipsoid plus an external shear. Then, we measure
 the ratio of fluxes in some apertures at the source plane where the lensed
 images are inversely mapped. We find that the aperture flux ratios of band 7
 continuum image are perturbed by 10-20 percent with a significance at 
$2\,\sigma \sim 3\, \sigma$ level. Moreover,
we measure the astrometric shifts of multiply lensed images near
 the caustic using the CO(8-7) line. Using a lens model best-fitted to
 the band 7 continuum image, we
 reconstruct the source image of the CO(8-7) line by taking 
linear combination of inverted
 quadruply lensed images. At the 50th channel (rest-frame velocity 
28.6 $\tr{km}\,\tr{s}^{-1}$) of the CO(8-7) line, we find an
 imprint of astrometric
 shifts of the order of $0.01$ arcsec in the source image. 
Based on a semi-analytic calculation, we find that the 
observed anomalous flux ratios and the astrometric shifts can be
explained by intergalactic dark structures in the line of sight. A compensated
 homogeneous spherical clump with a mean surface mass density of the
 order of $10^8\,\ms h^{-1}\tr{arcsec}^{-2}$ can explain the 
observed anomaly and astrometric shifts simultaneously. 
\end{abstract}

\section{Introduction}
The nature of matter distribution on sub-galactic scales
has been veiled in mystery.
In comparison with the prediction from $N$-body simulations, 
the observed 
density profiles of dwarf galaxies are not cuspy but cored \citep{navarro1996, moore1999, swaters2003,
simon2005}, the number of observed satellite galaxies in our Galaxy is 
too small \citep{klypin1999,moore1999}, and the observed circular
velocities of most massive subhaloes in our Galaxy are too small \citep{boylan-kolchin2011, wang2012}.   

Gravitational lensing provides a powerful tool to measure
the matter distribution on sub-galactic scales. 
It has been known that 
some quadruply lensed quasars show anomalies in the observed 
flux ratios of lensed images assuming that  
the gravitational potential of the lens is
sufficiently smooth. Such a discrepancy is 
called the ``anomalous flux ratio'' and 
has been considered as an imprint of cold dark matter subhaloes
with a mass of $\sim 10^{8-9} \ms$ in lensing
galaxies \citep{mao1998,metcalf2001,chiba2002,dalal-kochanek2002,
keeton2003, metcalf2004,chiba2005,inoue-chiba2005a,inoue-chiba2005b, 
sugai2007,mckean2007,more2009,minezaki2009, xu2009, xu2010, vegetti2012}. 

However, intergalactic haloes in the line of sight 
can act as perturbers as well \citep{chen2003,metcalf2005a,xu2012}.
Indeed, taking into account the astrometric shifts of lensed images, 
recent studies have shown that the observed anomalous flux ratios can be 
explained solely by intergalactic structures that consist of haloes,
filaments and voids with a surface mass density $\sim 
10^{7-8}\, \ms h^{-1} \textrm{arcsec}^{-2}$ \citep{inoue-takahashi2012,
takahashi-inoue2014,inoue2015,inoue-etal2015} without considering
subhaloes in the lens galaxy. 

In order to determine the origin of flux-ratio anomalies, we need to
measure the flux ratios and positions of lensed images 
as precisely and as many as possible. If we find that 
the probability of having flux-ratio anomalies increases
as a function of the source redshift, then it will be a strong indication 
that the anomaly is caused by the line-of-sight structures rather than
subhaloes. For achieving this scientific goal, strongly 
lensed submillimetre galaxies (SMGs)
are ideal targets as their redshifts are biased at $z=$2-3 \citep{simpson2014}. 

The lens system SDP.81, also known as H-ATLAS J090311.6+003906 
is one of such systems. The source is an SMG at
$z=3.042$ \citep{negrello2010} and the primary lens is a massive
elliptical galaxy at $z=0.2999$ \citep{negrello2014}. 
The property of the source has been extensively studied \citep{rybak2015-a,rybak2015-b,tamura2015,hatsukade2015,wong2015}.
However, we note that all of these models are based on  
smooth potentials without fully incorporating the small-scale structure of 
the lens system.

In this paper, we report our analysis of 
the ALMA long baseline science verification data 
of SDP.81 (used by the authors previously mentioned) 
to investigate possible gravitational perturbations by dark
(sub-)structures along the line of sight to the lens system.

This paper is organized as follows. In section 2, we describe the data.
In section 3, we explain the model of the primary lens and our method for
reconstructing and modelling the data. In section 4, we present
our results on the fitting of our model. In section 5, we show the 
semi-analytic estimate of perturbation by intergalactic dark structures
and the analytic estimate of possible perturbation by a spherically
symmetric homogeneous clump. In section 6, we conclude and discuss 
some relevant issues.

In what follows, we assume a cosmology 
with matter density $\Omega_{m,0}=0.3134$, baryon density 
$\Omega_{b,0}=0.0487$, a cosmological constant $\Omega_{\Lambda,0}=0.6866$,
a Hubble constant $H_0=67.3\, \textrm{km}\,\textrm{s}^{-1} \textrm{Mpc}^{-1}$,
a spectrum index $n_s=0.9603$, and the root-mean-square (rms) 
amplitude of matter fluctuations at $8 h^{-1}\, \textrm{Mpc}$, 
$\sigma_8=0.8421$, which are obtained from the observed 
CMB (Planck+WMAP polarization; \citet{ade2014}).
In plots of images, the horizontal and vertical coordinates correspond  
to R.A and Dec. in arcsec, respectively. 

\section{DATA}
We used the ALMA science verification data on SDP.81 taken from 
the ALMA Science Portal (see \citet{vlahakis2015} for detail). 
SDP.81 was observed in 2014 October as part of
the Long Baseline Campaign at band 4(151 GHz), 6(235 GHz) and 7(290 GHz). 
In addition to continuum, the bands 4, 6, and 7 data include lines of
CO(5-4)(observed frequency=142.570 GHz), CO(8-7)(228.055 GHz), and CO(10-9)(285.004 GHz), 
respectively. We used the processed archival images of
the band 6 and 7 continuums and the CO(8-7) line, whose calibration 
and imaging were carried out using the Common Astronomy Software
Applications (CASA) \citep{CASA}. The images were processed using the CLEAN algorithm
with a $robust=1$ weighting (Briggs weighting) of the visibilities.
The CO data were binned spectrally into channels with 
21.0 $\textrm{km}\, \textrm{s}^{-1}$ wide. In this paper, we used the 
band 6 and 7 continuum images without $uv$-tapering
and the CO(8-7) images with $uv$-tapering (1000k$\lambda$) to a
resolution of $\sim 170\,$mas in order to increase the signal-to-noise
ratio on each pixel. In order to investigate astrometric shifts, we 
chose CO(8-7) images at channels 47 to 52 (rest-frame velocities 
from -34.4 $\tr{km}\,\tr{s}^{-1}$ to 70.6 $\tr{km}\,\tr{s}^{-1}$). 
The pixel sizes are $0.005$ and $0.02$ arcsec, respectively. The
semi-major and semi-minor axes and the position angles (PAs) of the 
synthesized beams are $31\times 23\,$mas (PA$=16^\circ$)
and $169 \times 117\,$mas (PA$=47^\circ$), respectively.

\section{Method}
\subsection{Canonical Lens Model}
As a canonical model of the unperturbed lensing galaxy, 
we adopt a singular isothermal ellipsoid 
(SIE) \citep{kormann1994}, which has been widely used in lens modelling 
and has successfully reproduced many other lens systems (e.g., 
\citet{keeton1998}). In order to fit the model, 
we used the relative positions of lensed quadruple
images and the centroid of lensing galaxy obtained from
CLEANed images. As is well known, the CLEAN algorithm
is nonlinear and produces real-space images 
with non-trivial noise properties. As a result 
it is not possible to easily come up with a statistically
well justified error when the lens modelling is done.  
In order to minimize this effect, we only used positions 
of brightest peaks with a signal-to-noise ratio S/N$\gtrsim 10$.
The other fainter spots along the Einstein ring are more prone
to such noises. The contribution from groups, clusters, and large-scale
structures at angular scales larger than the
Einstein angular radius of the primary lens was taken into account as an external
shear (ES). The parameters of an SIE plus an ES (SIEES) model 
are the effective Einstein angular radius (the 
mass scale inside the critical
curve) $b$\footnote{We use the definition adopted in \citet{kormann1994}.}, the apparent ellipticity $e$ of the lens and its
position angle $\theta_e$, the strength and the direction of the external shear
$(\gamma,\theta_\gamma)$, the lens position $(x_{G},y_{G})$, and the source
position $(x_s,y_s)$. The angles $\theta_e$ and $\theta_\gamma$ are
measured counter-clockwise from North. 

In order to implement the simultaneous $\chi^2$ fitting of the positions
of the centroids of lensed images and the lensing galaxy, we used our
developed code. We checked the accuracy of our code  
by comparing it with other numerical codes such as GRAVLENS \footnote{See
http://redfive.rutgers.edu/$\sim$keeton/gravlens/}. 
The obtained $\chi^2$ values are consistent each other within a few percent.

\subsection{Source Reconstruction}
If a lens model is perfectly correct and observational 
noises are negligible, the source images 
that are inverted from multiple images should be all identical. However, in
practical setting, they are different to some extent 
due to the error in the model, the noise and the finite resolution 
in the observed image. In order
to reconstruct the brightness distribution of a source for a given lens
model, we use linear combination of source images that are 
inverted from multiple images.
 
Let $c_i$'s and $f_i$'s $(i=1,2,\cdots,N)$ 
be arbitrary real functions that represent ``weights'' and surface brightness
in the source plane corresponding to $N$-multiply
lensed images. The true surface brightness $f_{true}$ of a 
source can be estimated by a linear combination
\BE
f_{true}\sim f(\v{r};\v{c})\equiv \f{\sum_{i=1}^{N}c_i(\v{r}) f_i(\v{r}) }{\sum_{i=1}^{N} c_i(\v{r})},
\EE
where $\v{r}$ is the position in the 
proper coordinates in the source plane. In
principle, the weight function $c_i$ can be determined by changing 
weights at each pixel in the source plane. 
However, the degree of freedom of weights is $N$ times
larger than that of the $f_{true}$. Therefore, we have to put a certain
prior on the weights. In what follows, 
we use the following two types of weighting scheme; 
the ``uniform weight'' (U-weight) in which $c_i\equiv 1$ and the
``magnification weight'' (M-weight) in which $c_i=\mu_i$ where $\mu_i$ is
the magnification for the $i$-th lensed image.
The uniform weight is preferable in the cases where the effect of
astrometric shifts by subhaloes or intergalactic structures dominates over 
the observational noise effect. As this scheme puts the weight equally
over the inverted images, the position errors for the best-fitted model with a smooth potential 
are minimized provided that the perturbation scale is sufficiently
smaller than the Einstein radius of the primary lens. 
In contrast, the magnification weight is 
preferable in the cases where the observational noise effect dominates 
over the effect of perturbation. In fact, it corresponds to the 
inverse-variance weighted average if the observational noise in the
image plane is homogeneous.

\subsection{Modelling}
In this analysis, we used only the positions of bright 
clumps that are not too close to the caustic 
and a possible active galactic nucleus (AGN) emission in the band 7 continuum image
\citep{vlahakis2015,tamura2015} for modelling the unperturbed lens. Note
that the angular resolution of the band 7 data is the highest in the data set.
Our modelling procedure can be verified as follows. 
First, the expected change in the flux ratios are of the order of 10
percent \citep{inoue-takahashi2012}. 
Therefore, it is difficult to measure such a tiny change with 
much fainter clumps in the observed arc. Second, bright clumps that are
too close to the caustic are sensitive to astrometric shifts caused by
subhaloes or intergalactic structures. Typical astrometric shifts due to subhaloes or intergalactic 
structures are of the order of $0.01$ arcsec \citep{inoue-takahashi2012}. 
These shifts yield significant changes in the fluxes of these 
clumps and consequently make a lens model with a smooth potential difficult to fit. 
 
\begin{figure}
\hspace{-0.16cm}
\vspace{-1.0cm}
\IG[width=90mm]{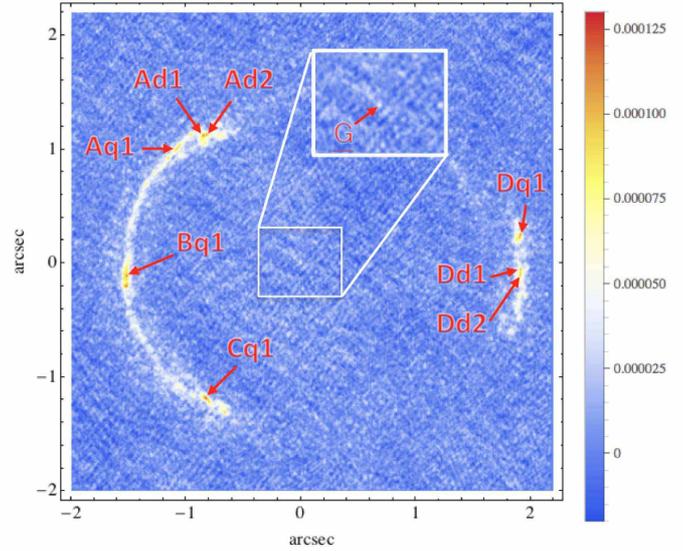}
\vspace{0.3cm}
\caption{Band 7 continuum image of SDP.81. 
The unit is Jy per beam. The quadruply lensed clumps
are Aq1, Bq1, Cq1, and Dq1. The doubly lensed ``north'' and ``south'' clumps
are Ad1, Dd1 and Ad2, Dd2, respectively. G is a possible AGN emission.
~~~~~~~~~ }
\label{band 7}
\end{figure}

\begin{figure}
\hspace{-0.2cm}
\vspace{-0.4cm}
\IG[width=92mm]{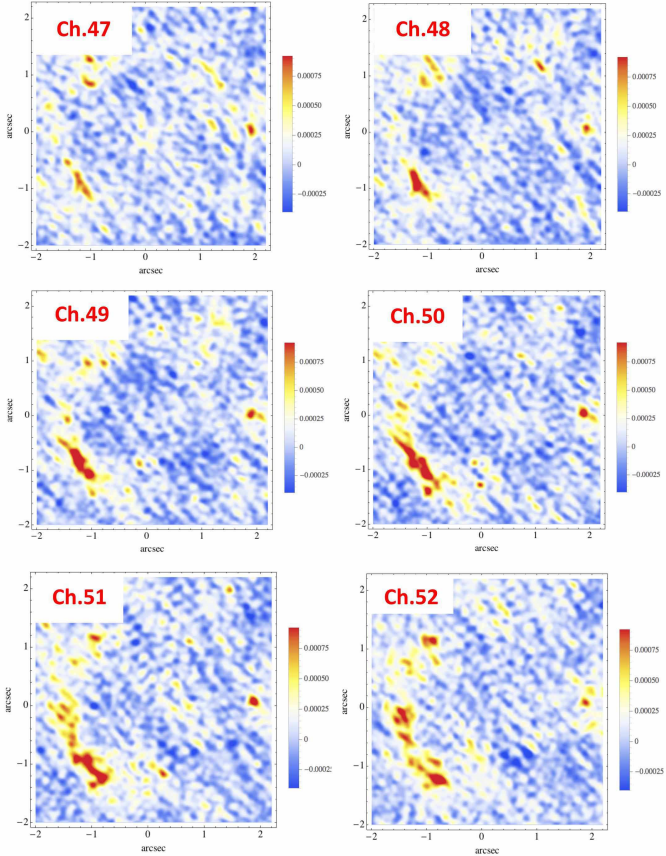}
\caption{CO(8-7) line images of SDP.81 at channels 47 to 52 (rest-frame velocities 
from -34.4 $\tr{km}\,\tr{s}^{-1}$ to 70.6 $\tr{km}\,\tr{s}^{-1}$). 
The unit is Jy per beam. }
\label{CO(8-7)}
\end{figure}

After a careful analysis of both the band 7 continuum (Fig. \ref{band 7}) and the CO(8-7)
line images (Fig. \ref{CO(8-7)}), we found that the
band 7 image can be well fitted by a fold-caustic lens with one set of quadruples
(Aq1,Bq1,Cq1,Dq1) and two sets of doubles (Ad1,Ad2 and Dd1,Dd2) as demonstrated
in Fig. \ref{band 7}. B and D have a positive parity and A and C have a negative parity. 
In order to fit the model, we used the positions of the brightest peaks of
lensed images of these three sets (clumps) except 
for Aq1, which is difficult to identify.

As the signal to noise ratios of these clumps are
$\gtrsim 10$ on each pixel, the errors in the relative 
positions of the clumps are expected to be 
much smaller than the band 7 beam size $23 \times 31$
mas. If the clump consists of a point source and the observational noise is dominated
by thermal noise, we expect that the position measurement uncertainty
for the peaks is given by 
\BE
\Delta \theta \approx
\biggl(\frac{4}{\pi}\biggr)^{1/4}\frac{1}{\sqrt{8 \ln{2}}}\frac{\theta_{\tr{beam}}}{SNR}\approx 
0.451\frac{\theta_{\tr{beam}}}{SNR},
\label{accuracy}
\EE
where $\theta_{\tr{beam}}$ is the synthesized beam size (measured in
FWHM) and $SNR$ is the signal-to-noise ratio \citep{reid1988, honma2014},
which yields $\Delta \theta =1 -  1.4\, \tr{mas}$\footnote{Note that the 
nominal astrometric accuracy from the ALMA Long Baseline Campaign tests is
an rms positional error of 1.5\,mas, which is for an average
calibrator-target separation of 6 deg with an observing period
of one hour with a maximum baseline of 12\,km\citep{fomalont2015}.}. In
the model fitting, taking account of possible complex structures of the
source within the beam, we used more conservative values for the position measurement
uncertainty: the 1$\,\sigma$ errors are 1/4 of $\theta_{\tr{beam}}$, i.e., $6 \times 8$\,mas with the same PA
direction. We also used the position of the centroid of image G, a possible
emitting region from the AGN in the lensing galaxy. G was identified
by overlaying the continuum image in the band 6 with that in the band 7.
The distance between the centroids of G in the band 6 and 7 is $0.02$\, arcsec.  
Therefore, we assumed that the 1$\,\sigma$ error in the position of the centroid of 
G is $0.02$\,arcsec, which is similar to the beam size $\theta_{\tr{beam}}$. 

In what follows, we call a quadruply lensed image 
that belongs to the inverted image of the region inside the caustic and 
contains Aq1, image A. Images B, C, and D
are defined in a similar manner. The centre $(0,0)$ of the coordinates in the
image plane $(x,y)$ is set at the centroid of image G. 

\begin{figure*}
\hspace{-0.2cm}
\IG[width=170mm]{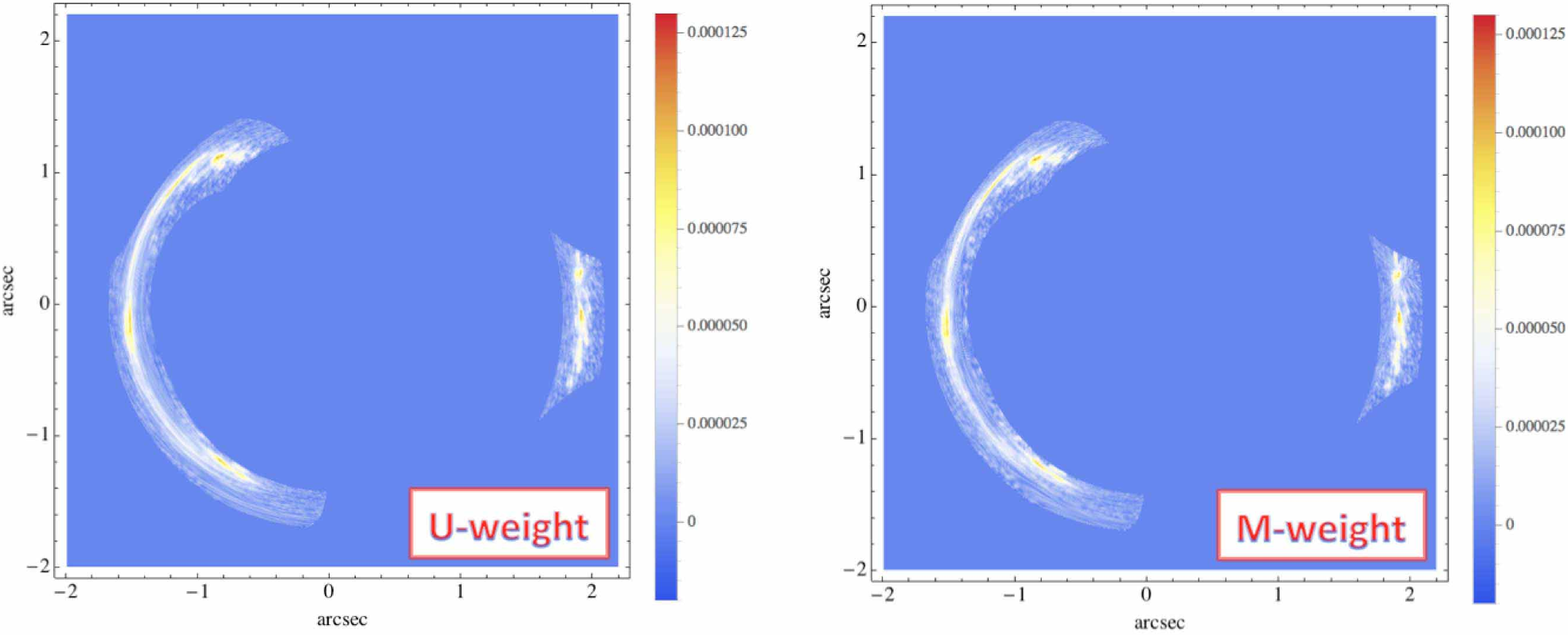}
\caption{Modelled surface brightness distribution in the image plane for the uniform
weight (left) and the magnification weight (right) for the best-fitted
 SIEES model based on the band 7 continuum image. The unit is Jy per beam.  }
\label{SIEES-lens}
 \end{figure*}

\begin{figure*}
\hspace{-0.36cm}
\IG[width=180mm]{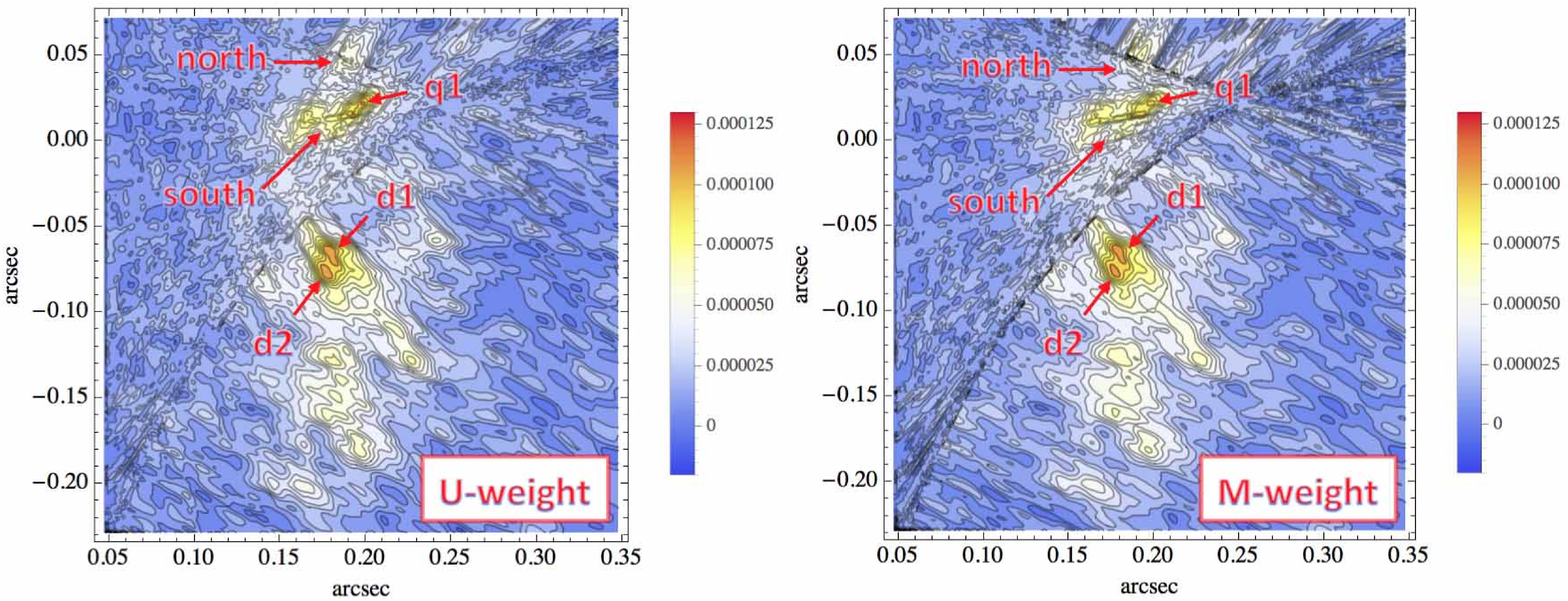}
\caption{Modelled surface brightness distribution in the source plane for the uniform
weight (left) and the magnification weight (right) for the best-fitted
 SIEES model based on the band 7 continuum image. The unit is Jy per beam.}
\label{SIEES-source}
\end{figure*}

\begin{figure}
\hspace{-0.16cm}
\IG[width=85mm]{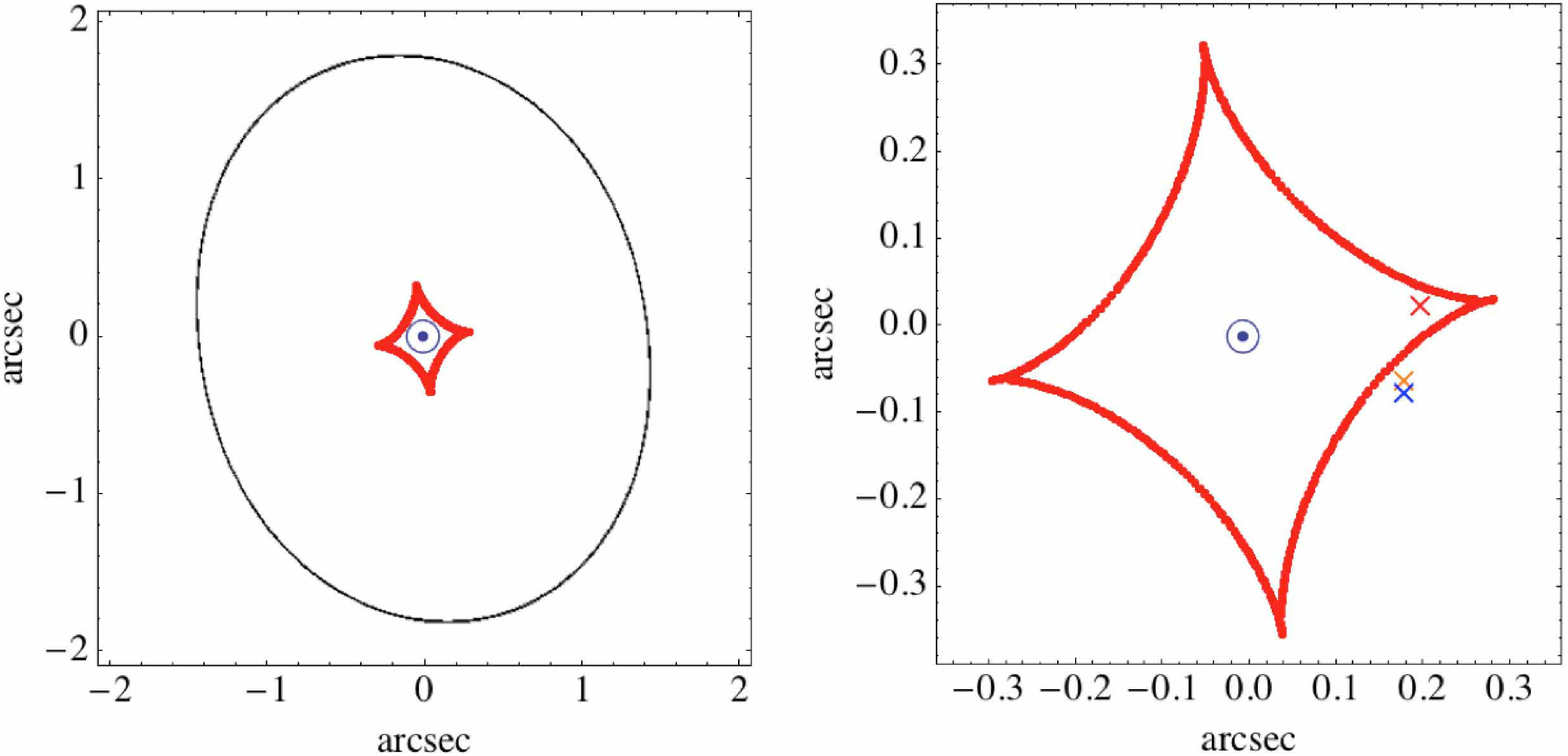}
\caption{Critical curve (black) and caustic (red) for the best-fitted
 SIEES model based on the band 7 continuum image. The centre of the coordinates is at image G. The circled dot
 indicates the position of an SIE and crosses represent the positions of
 bright clumps.}
\label{SIEES-caustic}
 \end{figure}

\section{Result}
\subsection{Best-fitted Model}
The parameters of the best-fitted SIEES model based on the band 7
continuum image is shown in Table
\ref{table-SIEES}. The degree of freedom for the observables is 16 and
that for the model parameters is 13. Therefore, the reduced $\chi^2$ 
($=\chi^2$/dof) is $0.647/(16-13)=0.22$. The mean distance 
between the position of the best-fitted and
observed lensed images was found to be $0.0014$\,arcsec. 
This over-fitting may imply 
that the estimate of position accuracy (=beam
size/4) in our analysis is a conservative one and astrometric shifts
by perturbers are sufficiently small for the quadruples of q1.  

The lensed images, the caustic and critical curve, and the reconstructed source images 
for the best-fitted model are shown in 
Figs. \ref{SIEES-lens}, \ref{SIEES-source}, and \ref{SIEES-caustic}, respectively. 
The 1\,$\sigma$ uncertainty in the model parameters are obtained by
finding a range of each parameter that gives 
the reduced $\chi^2$ less than 1. 

Our result is in agreement with the parameters obtained by
\citet{rybak2015-a}, \citet{dye2015} and \citet{tamura2015}. Though our
best-fitted parameters are slightly deviated from their values, the differences
fall within 1$\,\sigma$-2$\,\sigma$ errors in our analysis. 

Apart from minor differences in the lens model (power-law index, core, etc.), 
the slight differences in the best-fitted parameters mainly come from degeneracy between 
ellipticity $e$ and external shear $\gamma$: the sum of 
$e$ and $\gamma$ is nearly constant. In order to investigate this effect,
we searched for another set of parameters that are much closer to those
in \citet{dye2015} by changing $e$ and $\gamma$. Then we found a set of ``concordant'' model parameters 
(Table \ref{table-SIEES-con}). The differences between the
``concordant'' parameters and those in \citet{dye2015}, which uses a
semi-linear inversion method, fall within $~1\,\sigma$ errors in their data. 
The reduced $\chi^2$ ($=\chi^2$/dof) for the ``concordant'' model is $1.51/(16-13)=0.50$. 

Moreover, we also computed parameters for which the orientations of the
SIE and an external shear $\theta_e, \theta_\gamma$ coincide with 
the values in \citet{dye2015} as well
as $e$ and $\gamma$ for checking consistency
(Table \ref{table-SIEES-semilinear}). 
We fixed the power-law index $\alpha=2$ of the
gravitational potential of the primary lens, 
and optimized the parameters of $b$ and the source positions $(x_s, y_s)$ 
of the three clumps. The reduced $\chi^2$ ($=\chi^2$/dof) 
for the model based on a semi-linear inversion method is
$5.18/(16-13)=1.7$. Thus, our best-fitted values are slightly
deviated from those in \citet{dye2015} but the difference falls
within $2\,\sigma$.

%Errors from ``sdp81NN-lensmodel-SIEES-3cores-model4N-newc-error.nb'' 
\begin{table*}
\caption{Best-fitted Model Parameters}
\begin{tabular}{ccccccc}
\hline
$\chi^2/\textrm{dof}$ &  ($x_\tr{G},y_\tr{G}$)(arcsec) & $b$ (arcsec) & $e$ &
   $\theta_e$ (deg) & $\gamma$ & $\theta_\gamma$ (deg) \\ 
\hline
$0.647/3$ & $(-0.007, -0.017)$ & $1.605^{+0.030}_{-0.004}$ & $0.17^{+0.17}_{-0.09}$  
& $25^{+44}_{-14}$ & $0.057^{+0.038}_{-0.039} $ &
			 $-10^{+10}_{-0}$ \\
\\ 
\hline
 & ($x_\tr{s},y_\tr{s}$)(arcsec) for q1 &  ($x_\tr{s},y_\tr{s}$)(arcsec)
     for d1    & ($x_\tr{s},y_\tr{s}$)(arcsec) for d2 & & &  \\ 
\hline
 & $(0.1978,0.0215)$ & $(0.1783, -0.0645)$ & $(0.1781,-0.0780)$  & & & 
\\
\label{table-SIEES}
\end{tabular}
\medskip
\end{table*}

\begin{table*}
\caption{Concordant Model Parameters}
\begin{tabular}{ccccccc}
\hline
$\chi^2/\textrm{dof}$ &  ($x_\tr{G},y_\tr{G}$)(arcsec) & $b$ (arcsec) & $e$ &
   $\theta_e$ (deg) & $\gamma$ & $\theta_\gamma$ (deg) \\ 
\hline
$1.51/3$ & $(-0.024, -0.005)$ & $1.609 $ & $0.20  $ & $16 $ & $0.041 $ &
			 $-8 $ \\
\\ 
\hline
 & ($x_\tr{s},y_\tr{s}$)(arcsec) for q1 &  ($x_\tr{s},y_\tr{s}$)(arcsec)
     for d1    & ($x_\tr{s},y_\tr{s}$)(arcsec) for d2 & & &  \\ 
\hline
 & $(0.1961,0.0312)$ & $(0.1862, -0.0579)$ & $(0.1860,-0.0722)$  & & & 
\\
\label{table-SIEES-con}
\end{tabular}
\medskip
\end{table*}
\begin{table*}
\caption{Model Parameters Based on Semi-linear Inversion \citep{dye2015}}
\begin{tabular}{ccccccc}
\hline
$\chi^2/\textrm{dof}$ &  ($x_\tr{G},y_\tr{G}$)(arcsec) & $b$ (arcsec) & $e$ &
   $\theta_e$ (deg) & $\gamma$ & $\theta_\gamma$ (deg) \\ 
\hline
$5.18/3$ & $(-0.033, 0.005)$ & $1.606 $ & $0.20  $ & $13 $ & $0.040 $ &
			 $-4 $ \\
\\ 
\hline
 & ($x_\tr{s},y_\tr{s}$)(arcsec) for q1 &  ($x_\tr{s},y_\tr{s}$)(arcsec)
     for d1    & ($x_\tr{s},y_\tr{s}$)(arcsec) for d2 & & &  \\ 
\hline
 & $(0.1904,0.0375)$ & $(0.1873, -0.0515)$ & $(0.1872,-0.0664)$  & & & 
\\
\label{table-SIEES-semilinear}
\end{tabular}
\medskip
\end{table*}

\subsection{Flux in Aperture}
If a lensed image is significantly
distorted, it is difficult to correctly measure the 
fluxes of an identical component of the source in the 
image plane. Therefore, we measure
the relative aperture fluxes of quadruply lensed 
images at the \ti{source plane}. If the model
is perfect and the noise is negligible, 
then the inverted aperture flux ratios should be equal to 1.
Any deviation from 1 indicates anomaly in the flux ratio,
unless it is caused by observational errors.

\begin{figure*}
\hspace{0.00cm}
 \IG[width=175mm]{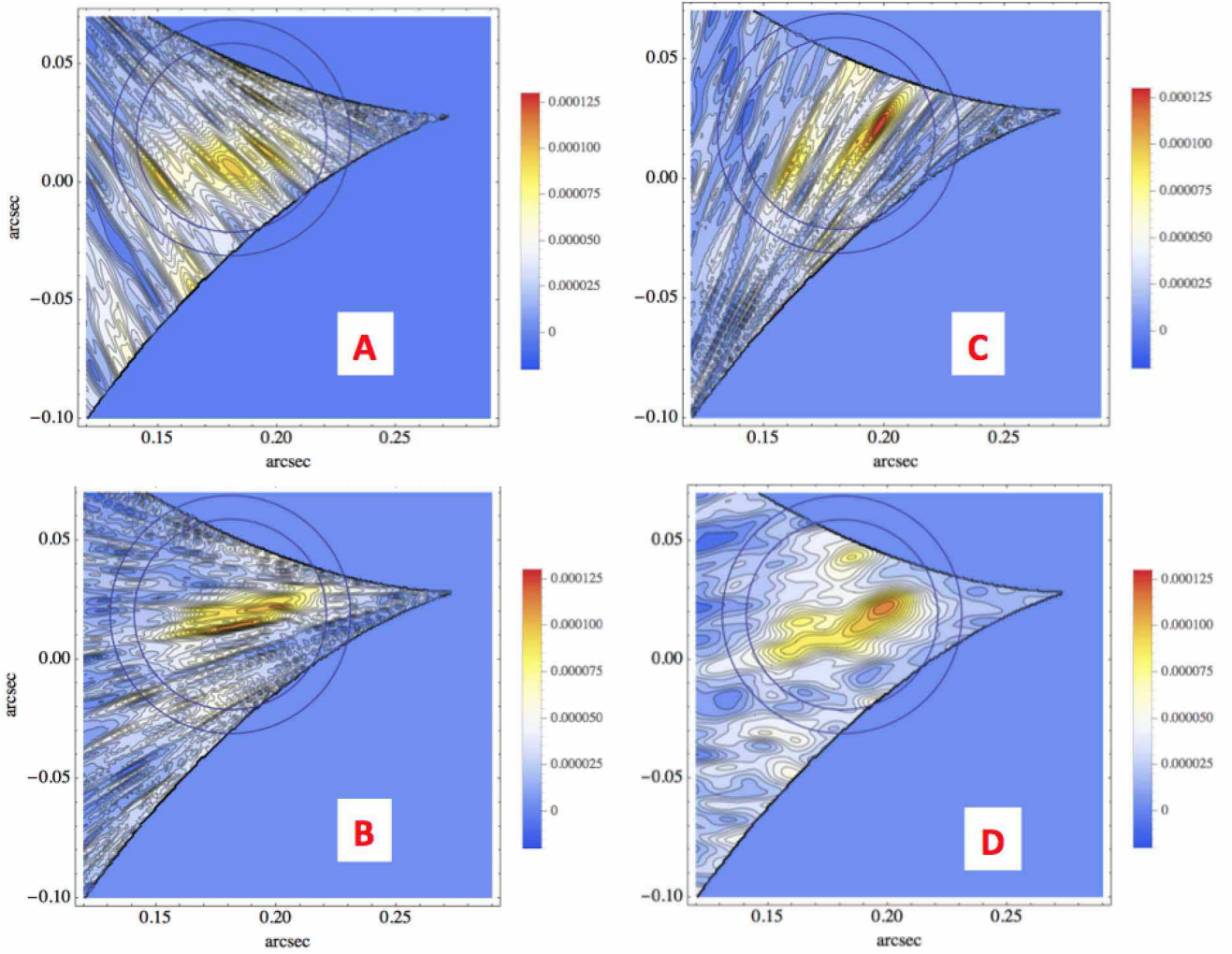}
\caption{Inverted quadruple images of band 7 data using a best-fitted SIEES
 model. The unit is Jy per beam. Blue circles denote apertures with a
 radius $0.04$ and $0.05$ arcsec. The centre of the apertures is 
$(0.1809, 0.01882)$. }
\label{SIEES-inv}
\end{figure*}

In order to invert the observed image in the image plane back to the
source plane, we use three models, namely the ``best-fitted'', the most
probable model, the ``concordant'', which is concordant with models in 
literature, and the ``semi-linear inversion'' model introduced in
\citet{dye2015}. In order to reduce the systematic effect of beam
smoothing, the size of an aperture radius is taken to be sufficiently larger
than the inverted beam size while most of the flux is
contained inside the choice of aperture (see Fig. \ref{SIEES-inv}). 

As shown in Fig. \ref{SIEES-source}, 
the quadruply lensed source consists of two ``north'' and ``south'' 
extended components. A component $\rm{q}1$ that yield quadruples in the source plane is 
at the right end of the ``south'' region which may have more complex
structures. The centre of apertures
was chosen so that these components are contained within a radius of $0.04$
arcsec. As the maximum linear sizes of the inverted beams are 
less than $\sim 0.04$ arcsec, an aperture radius of $>0.04$ arcsec
is necessary to reduce the beam smoothing effect.  

To estimate the significance of deviation in the ratios of fluxes in 
aperture, we have carried out Monte Carlo simulations using subsamples
in the data.  
First, we randomly selected
100 points within a ring region with an angular distance $R$ from the centre of
the lens satisfying $3.5<R<4.3$ arcsec. The region with $R<3.5$ arcsec was not used
because we observed a systematic deviation from zero in the fluxes. 
It may be caused by either a part of broad Einstein ring of lensed
image or possible foreground emission from the lensing galaxy. 
Thus the observed fluxes in this ring region are expected to be dominated by  
observational noises. Then for each point, we made a translation of the 
image of the diamond-shaped region inside the caustic for each multiple
image separately so that the lensed images of $q1$ coincide with the
point. Finally, we inverted the observed band 7 image within the
translated region back to the source plane and computed the fluxes in
circles with aperture radius $0.04,0.05,0.06$ arcsec to obtain
observational errors in the source plane.    

The leak of flux outside or inside aperture can be a source of systematics. In
order to estimate the effect, we carried out a hypothetical observation 
in which the source image is identical to the one reconstructed from
image A and image D using the 
magnification weight. Then we convolved the obtained image
with an elliptic Gaussian beam that corresponds to the one in the band 7
continuum image and computed aperture fluxes. We have found that the
errors in the aperture flux ratios are less than 4 percent. As the reconstructed lensed images of the tentative source image are smoothed twice, the
actual size of the source clumps are expected to be much smaller. Thus
our test of systematics gives an upper limit on the error caused by the beam shape. 

In Table 4, we show our results of the aperture 
fluxes in the band 7 image. In this table, B/A, C/A and D/A represent aperture 
flux ratios of B to A, C to A, and D to A and ``deviation'' is defined as
1-(flux ratio) divided by $1\,\sigma$ error obtained from the randomly
chosen 100 samples. Note that the error caused by
the beam shape is not taken into account here. In the best-fitted model, 
the aperture fluxes of images B and C are deviated from 1 at $\sim 3\,\sigma$     
level whereas image D is equal to 1 within an error. In the concordant
model, the significance of anomaly in images B and C is lowered to $\sim
2 \,\sigma$, and image D is consistent with being 1. In the ``semi-linear inversion'' 
model, however, the significance of anomaly in C is further 
lowered by $\sim 1\, \sigma$ but that in B does not change.  

In the best-fitted model, 
the magnifications of the bright clump q1 
at images A and D are 6.87 and 3.92 and those 
at images B and C are 21.9 and 18.2.
As the perturbation of magnification divided by the 
unperturbed magnification is proportional to 
the unperturbed magnification \citep{inoue-takahashi2012}, 
the perturbation for image A is 2-3 times smaller than those for images B and
C. Perturbation for image D is much smaller. 
Thus, the expected perturbation due to line-of-sight
dark structures is much larger in images B and C than images A and D.
Therefore, the observed anomalous result for only images B
and C is consistent with the interpretation that they are 
perturbed by either subhaloes or the 
line-of-sight structures provided that the non-perturbed
smooth gravitational potential is described by that of an SIE and an ES. 

For the band 6 image, it turned out 
that the aperture flux-ratios B/A and C/A are
$\sim 0.9$ with slightly larger errors. Thus, we were 
not able to find any anomalous features in the aperture flux ratios in
the band 6 image. However, the obtained result is consistent with that for 
the band 7 image within $1\,\sigma$. 

The elongation of the inverted beams in the source
plane may cause some errors in estimating the aperture flux ratios. However, 
the deviation from 1 for D/A is less than 5 percent. Therefore, we expect that 
the beam effect is less than 5 percent provided that image A and image D
are not perturbed at all. This is consistent with the analysis using a
hypothetical observation that we have mentioned.
Thus, it does not change our conclusion 
that the aperture flux-ratios show anomaly at 
2\,$\sigma$-3\,$\sigma$ level. 

A strange feature is observed in B/A. If B is 
perturbed by a clump, B should be \textit{magnified} since B has a positive
parity \citep{inoue-chiba2005b, inoue-takahashi2012}. However, our result indicates that image B is \textit{demagnified} in
comparison with the prediction of a best-fitted model with a smooth potential.
The feature may be a problem if one tries to explain the anomaly
by putting a single subhalo in the lensing galaxy. However, it can be
easily explained by a line-of-sight dark structure with a negative density
region, which we discuss in the next section.
 
\begin{table*}
\begin{center}
\caption{Flux Ratios in Band 7}
  \begin{tabular}{c c c c c c c} 
\hline
  & aperture(arcsec) & $\hat{\eta}(1\sigma)$ & & B/A & C/A & D/A \\
\hline
  &  &  & ratio  & 0.820 & 0.801  & 0.945  \\ \cline{4-7}
  &  0.06 &$0.092(0.031)$ &   error & 0.062 & 0.060 & 0.073  \\ \cline{4-7}
  &   & & deviation & 2.9 & 3.4 & 0.8   \\ \cline{2-7}
  &   &  & ratio  & 0.829 & 0.815  & 0.973  \\ \cline{4-7}
 best-fitted &  0.05 &$0.092(0.031)$ &   error & 0.058 & 0.058 & 0.072  \\ \cline{4-7}
  &   & & deviation & 2.9 & 3.2 & 0.4   \\ \cline{2-7}
  &  &  & ratio  & 0.819 & 0.825  & 0.996  \\ \cline{4-7}
  &  0.04 &$0.097(0.030)$ &   error & 0.055 & 0.057 & 0.071  \\ \cline{4-7}
  &   & & deviation & 3.3 & 3.0 & 0.1   \\ \hline
\\
\\
\hline
  & aperture(arcsec) & $\hat{\eta}(1\sigma)$ & & B/A & C/A & D/A \\
\hline
  &  &  & ratio  & 0.857 & 0.868  & 0.982  \\ \cline{4-7}
  &  0.06 &$0.069(0.031)$ &   error & 0.064 & 0.063 & 0.076  \\ \cline{4-7}
  &   & & deviation & 2.2 & 2.1 & 0.2  \\ \cline{2-7}
  &   &  & ratio  & 0.857 & 0.877  & 0.991  \\ \cline{4-7}
 concordant &  0.05 &$0.069(0.029)$ &   error & 0.059 & 0.060 & 0.073  \\ \cline{4-7}
  &   & & deviation & 2.4 & 2.0 & 0.1   \\ \cline{2-7}
  &  &  & ratio  & 0.846 & 0.893  & 1.008  \\ \cline{4-7}
  &  0.04 &$0.073(0.025)$ &   error & 0.055 & 0.059 & 0.069  \\ \cline{4-7}
  &   & & deviation & 2.8 & 1.8 & 0.1   \\ \hline
\\
\\
\hline
  & aperture(arcsec) & $\hat{\eta}(1\sigma)$ & & B/A & C/A & D/A \\
\hline
  &  &  & ratio  & 0.868 & 0.919  & 1.01  \\ \cline{4-7}
  &  0.06 &$0.061(0.032)$ &   error & 0.068 & 0.068 & 0.080  \\ \cline{4-7}
  &   & & deviation & 2.0 & 1.2 & 0.1  \\ \cline{2-7}
  &   &  & ratio  & 0.856 & 0.913  & 1.01  \\ \cline{4-7}
 semi-linear &  0.05 &$0.067(0.030)$ &   error & 0.062 & 0.065 & 0.076  \\ \cline{4-7}
 inversion  &   & & deviation & 2.3 & 1.3 & 0.1   \\ \cline{2-7}
 \citep{dye2015} &  &  & ratio  & 0.843 & 0.927  & 1.05  \\ \cline{4-7}
  &  0.04 &$0.080(0.028)$ &   error & 0.057 & 0.063 & 0.072  \\ \cline{4-7}
  &   & & deviation & 2.8 & 1.2 & 0.6   \\ \hline
  \end{tabular}
  \end{center}
\label{table-band7-flux}
\end{table*}

\subsection{Astrometric Shift}
The perturbation of gravitational acceleration due to dark structures also
causes astrometric shifts of lensed images. This effect is usually 
difficult to observe as the observable signal of shifts is quite
subtle. However, if the 
source is very close to caustics, the effect becomes significant as
the magnification can be changed drastically by such shifts. ALMA is an
ideal tool for finding such an effect.

\begin{figure*}
\hspace{-0.16cm}
 \IG[width=170mm]{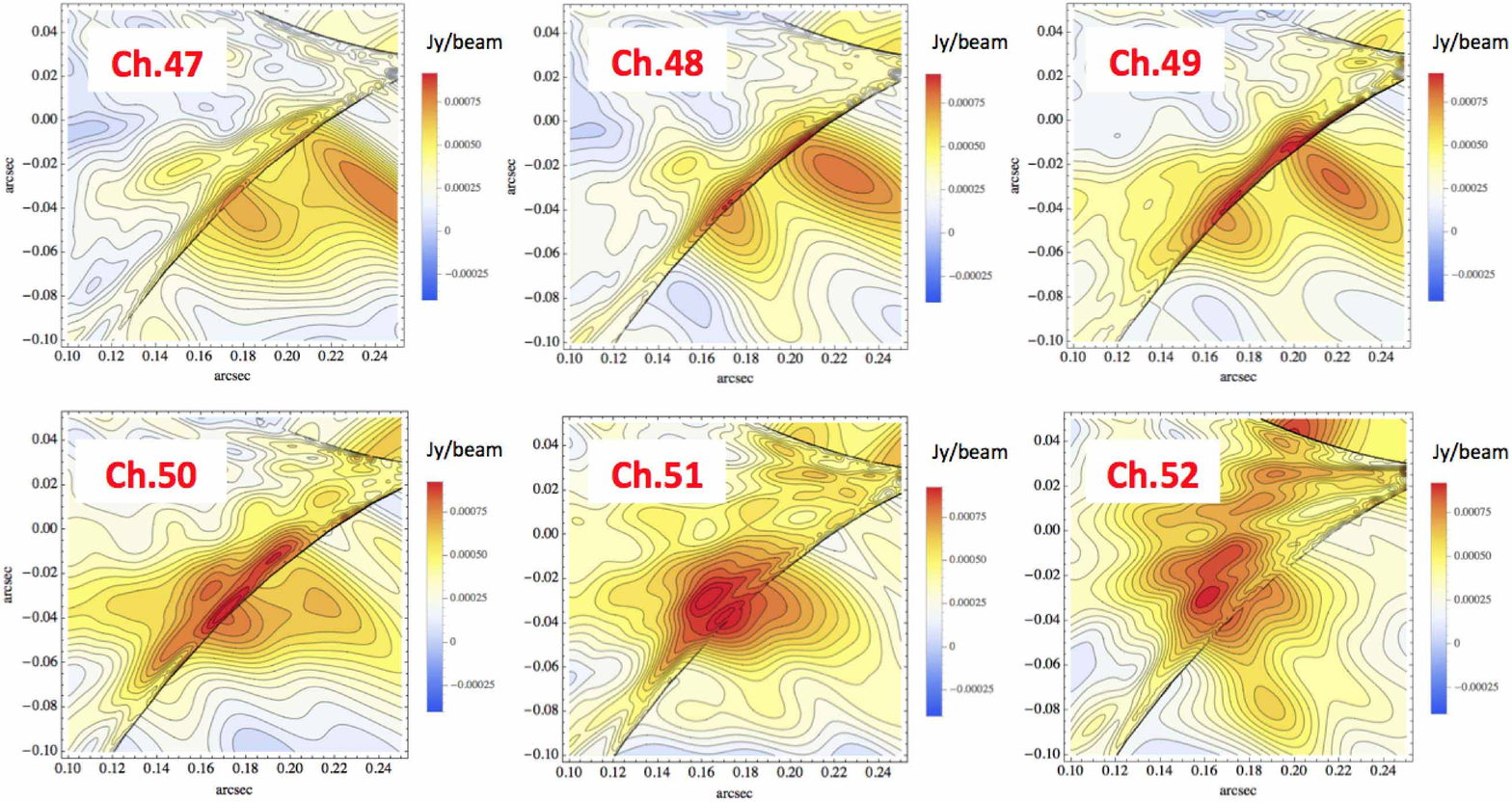}
\caption{Reconstructed images (uniform weight) of the CO(8-7) data (Ch.47 to
 Ch.52) using the best-fitted SIEES model. }
\label{CO-uweight}
\end{figure*}

\begin{figure*}
\hspace{-0.16cm}
 \IG[width=170mm]{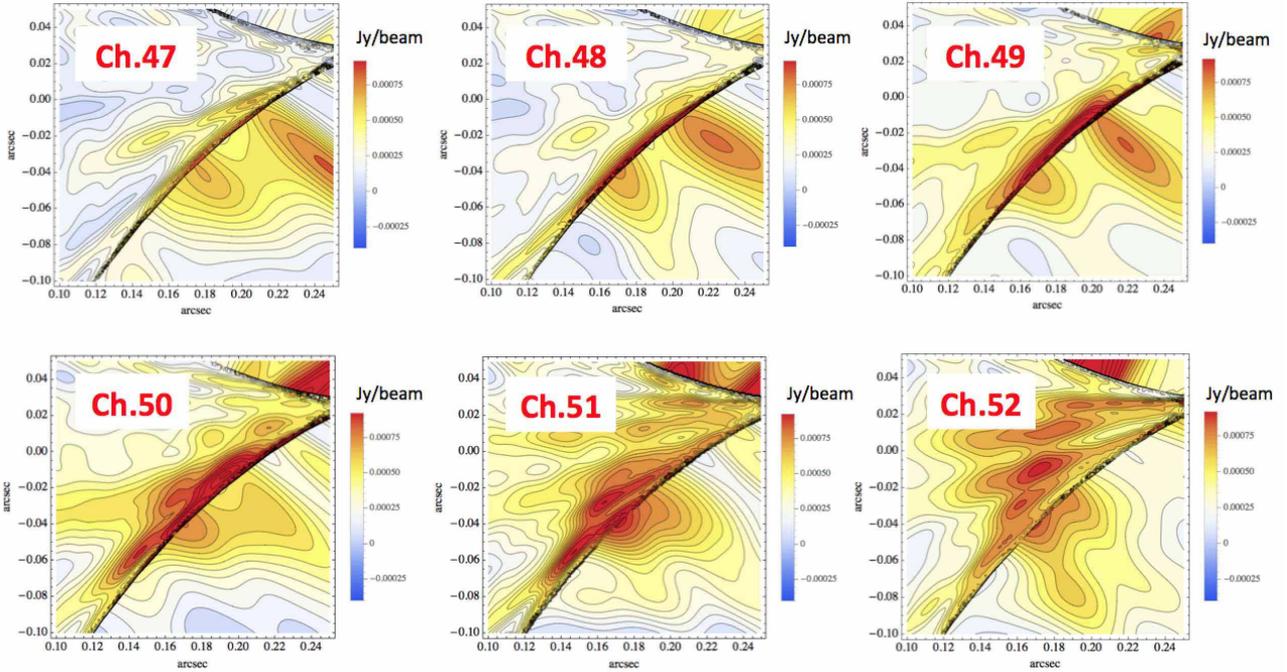}
\caption{Reconstructed images (magnification weight) of the CO(8-7) data (Ch.47 to
 Ch.52) using the best-fitted SIEES model. }
\label{CO-mweight}
\end{figure*}

\begin{figure*}
\hspace{-0.16cm}
 \IG[width=170mm]{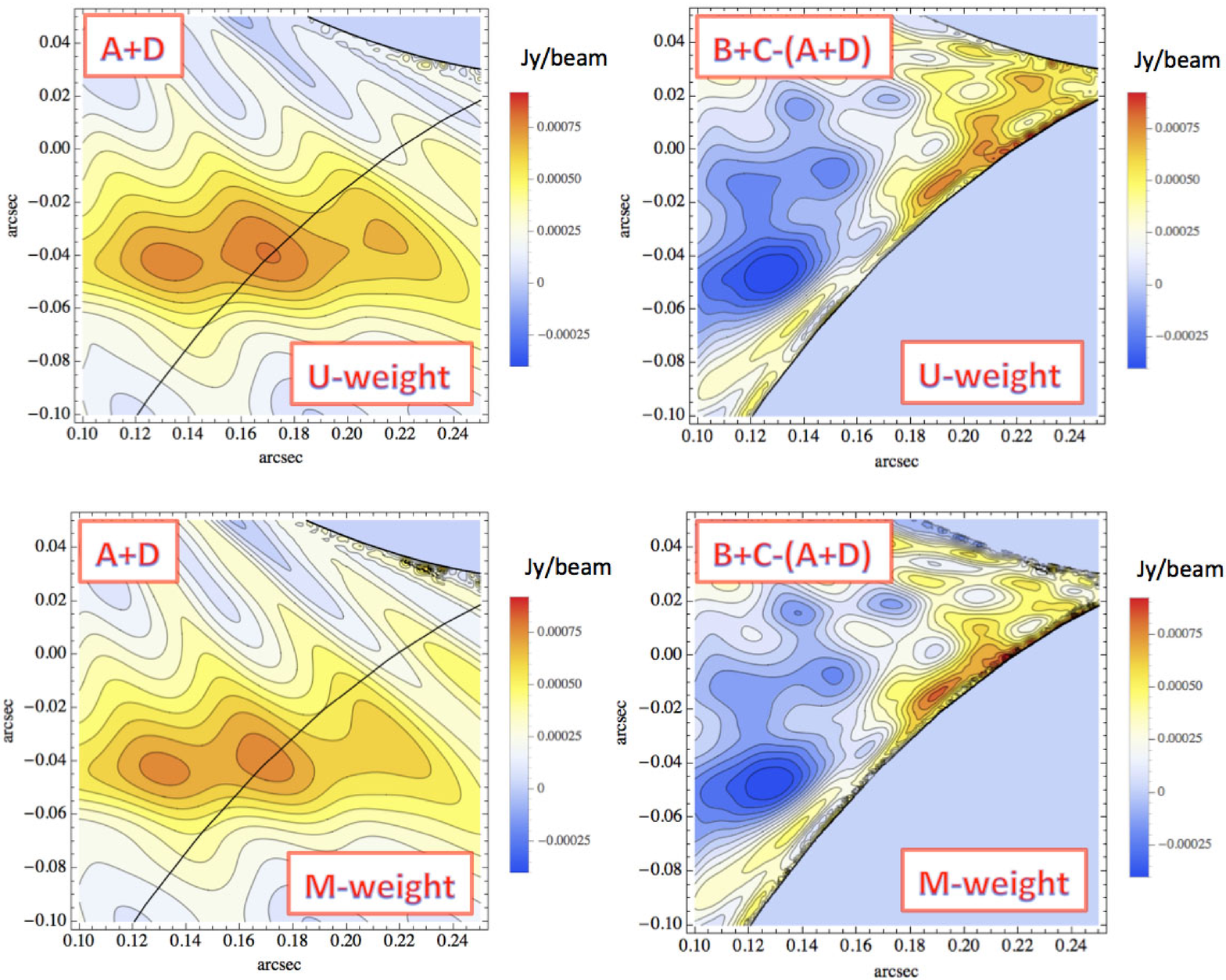}
\caption{Reconstructed images of the CO(8–7) Ch.50 data for the best-fitted SIEES model: image reconstructed from A and D images (left) and the difference
between an image reconstructed from B and C images and that
 reconstructed from A and D images (right).}
\label{CO-AD}
\end{figure*}

In order to find astrometric shifts, we analysed the inverted CO(8-7) line data of SDP.81
using the best-fitted lens model based on the band 7 data. As shown in 
Figs \ref{CO-uweight} and \ref{CO-mweight}, a discontinuous change
in the surface brightness across the caustic is observed at Ch.50. This suggests that a
bright clump is crossing the caustic at the Ch.50. In order to look into
the crossing, we plot the sum
of A and D images in Fig. \ref{CO-AD}.
One can clearly see that a bright clump centred at approximately  
$(0.18,-0.02)$ in the source plane is crossing the caustic, which
yields elongation of the inverted image along the caustic.
However, it turned out that the 
elongation is asymmetric around a bright clump. In order
to show this feature, we plot the difference in the surface brightness between
the inverted image reconstructed from B and C images (B+C) and that from
of A and D images (A+D) in Fig. \ref{CO-AD}. One can see a distinctive feature 
at the region centred at $\sim (0.19,-0.02)$ in the source plane. The
amplitude of the feature is $\gtrsim 3\,\sigma$. However,
no distinctive feature is observed at $\sim (0.16, -0.04)$. This is an
asymmetric feature around the brightest clump: the surface brightness 
changes discontinuously at the caustic.

This feature suggests astrometric shifts of the order of 0.01\,arcsec perpendicular to the
caustic for B and C images assuming that A and D images are not
perturbed at all. As the signal-to-noise ratio of the brightest 
clump near the caustic is found to be $\sim 8$, equation
(\ref{accuracy}) gives the expected positional uncertainty $\sim 0.01$\,arcsec.  
Therefore, the observed feature is much larger than the uncertainty.
If this is the case, B and C images might have been perturbed by some
dark structures in the line-of-sight. The observed anomalous feature in the aperture 
flux-ratios in band 7 data supports this interpretation.   

\section{Perturbation by Intergalactic Structures}
In this section, we briefly introduce the 
semi-analytic method for estimating the 
perturbative effect on the magnification
of lensed images caused by intergalactic 
structures in the line of sight. For detail, see 
\citet{inoue-takahashi2012, takahashi-inoue2014}.
\subsection{Semi-analytic Estimate}
We introduce a statistic $\eta$ adopted in \citep{inoue-takahashi2012} to measure the 
perturbation of magnification in strong lens systems. Let us consider 
a lens system with $N$-multiple images of a point source. The
magnification contrast of the image ''i'' is defined as $\delta^\mu_{\rm{i}} \equiv \delta \mu_{\rm{i}}/\mu_{\rm{i}}$,
where $\mu_{\rm{i}}$ is the magnification of the image ``i''
expected in the best-fitted model with a smooth gravitational potential
and $\delta \mu_{\rm{i}} $ is the perturbation of $\mu_{\rm{i}}$.
If the parity of an image is positive (negative), then the curvature of the arrival time surface 
at the image position is positive (negative). Therefore, the surface is locally
minimum or maximum (saddle) at the point. In what follows, we ignore 
images that correspond to maxima, since they are typically extremely faint. 

In terms of $\delta^\mu_{\rm{i}}$'s, the 
magnification perturbation $\eta$ is defined as
\BE
  \eta \equiv \biggl[\frac{1}{2 N_{\rm pair}} \sum_{{\rm i} \neq {\rm j}} \left[
 \delta^\mu_{\rm i} ({\rm minimum}) - \delta^\mu_{\rm j} ({\rm saddle})
 \right]^2 \biggr]^{1/2},
\label{eta_def}
\EE
where $ \delta^\mu_{\rm i} ({\rm minimum})$ and $ \delta^\mu_{\rm i} ({\rm saddle})$ are
the magnification contrasts of the minimum image and saddle image,
respectively, and $N_{\rm pair}$ is the total number of pairs of 
a saddle and a minimum images. The summation is performed 
over all the pairs. Roughly speaking, $\eta$ is equal to the mean value 
averaged over the all multiple images if correlations of flux perturbation 
between the lensed images are not taken into account. In SDP.81, there are two minima B and D and two saddles A and C. Then, we have
\BEA
 \eta &\equiv & \Biggl[\frac{1}{8} \biggl[\left(
 \delta^\mu_{\rm B} - \delta^\mu_{\rm A} 
 \right)^2+  \left(\delta^\mu_{\rm D} - \delta^\mu_{\rm A} 
 \right)^2+  \left(\delta^\mu_{\rm B} - \delta^\mu_{\rm C} 
 \right)^2
\nonumber
\\
&+& \left(\delta^\mu_{\rm D} - \delta^\mu_{\rm C} 
 \right)^2 \biggr] \Biggr]^{1/2}.
\label{eta_sdp81}
\EEA
Up to the linear order in $\delta_{\rm{i}}^\mu$, in terms of observed fluxes $\mu_{\tr A}, \mu_{\tr B},\mu_{\tr
 C},\mu_{\tr D}$ and estimated unperturbed fluxes 
 $\mu_{{\tr A}0}, \mu_{{\tr B}0},\mu_{{\tr C}0},\mu_{{\tr D}0}$, 
the estimator of $\eta$ is approximately given by
\BEA
 \hat{\eta} &\approx & \Biggl[\frac{1}{8} \biggl[\left(
 \frac{ \mu_{\tr B} \mu_{{\tr A}0} }{ \mu_{\tr A} \mu_{{\tr B}0}}-1 
 \right)^2+ \left(
 \frac{ \mu_{\tr D} \mu_{{\tr A}0} }{ \mu_{\tr A} \mu_{{\tr D}0}}-1 
 \right)^2
\nonumber
\\
&+& 
\left(
 \frac{ \mu_{\tr B} \mu_{{\tr C}0} }{ \mu_{\tr C} \mu_{{\tr B}0}}-1 
 \right)^2+ \left(
\frac{ \mu_{\tr D} \mu_{{\tr C}0} }{ \mu_{\tr C} \mu_{{\tr D}0}}-1 
 \right)^2 \biggr] \Biggr]^{1/2}.
\label{eta_est}
\EEA
For extended sources, we need to use aperture fluxes defined
at the source plane. If the source size is sufficiently small
in comparison with the Einstein radius of the primary lens,
and distance between the source and the caustic is sufficiently large, then
we can neglect the differential magnification effect. In that case, 
in terms of 
estimated unperturbed aperture fluxes 
 $m_{{\tr A}0}, m_{{\tr B}0},m_{{\tr C}0},m_{{\tr D}0}$ at the source plane,
the estimator of $\eta$ is approximately given by 
\BEA
 \hat{\eta} &\approx & \Biggl[\frac{1}{8} \biggl[\left(
 \frac{ m_{{\tr B}0} }{ m_{{\tr A}0}}-1 
 \right)^2+ \left(
 \frac{ m_{{\tr D}0}}{ m_{{\tr A}0}}-1 
 \right)^2
\nonumber
\\
&+& 
\left(
 \frac{ m_{{\tr B}0} }{ m_{{\tr C}0}}-1 
 \right)^2+ \left(
\frac{ m_{{\tr D}0} }{ m_{{\tr C}0}}-1 
 \right)^2 \biggr] \Biggr]^{1/2}.
\label{eta_app}
\EEA
In order to estimate the magnification perturbation 
$\eta$ theoretically,
we use a formalism based on the non-linear power spectrum
of density perturbation. 

First, we divide density fluctuations that cause 
lensing into two parts: strong lens and weak lens components.
The strong lens component corresponds to the primary lens
and the line-of-sight density fluctuations whose angular scales are 
comparable to or larger than 
the mean separation $\theta_0$ between the centre of the
primary lens and multiple images. In our case, it
corresponds to a best-fitted SIEES.  
The weak lens component
corresponds to the remaining line-of-sight density fluctuations whose angular
scales are smaller than the mean separation, and which causes astrometric shifts
and perturbation of fluxes. 

Secondly, we estimate the observed 
strength of weak lens component due to
line-of-sight structures using the residual
difference $\epsilon$ in the best-fitted positions of multiple images. 
The difference gives an upper limit of possible
contribution from line-of-sight structures. Since the contribution to
the magnification perturbation decreases as the comoving scale of the 
density fluctuations, we assume that the fluctuations with angular scales
just below $\theta_0$ are mostly affected. Thus, from two physical
scales $\theta_0$ and $\epsilon$, the second moment of $\eta$ can be estimated.   
In the Fourier space, these effects can be taken into account by 
considering a filtering function $W(k)$ to the non-linear power spectrum $P(k)$
obtained from $N$-body simulations.

For the bright clump q1 in the band 7 data, the mean
separation between the center of the lens and lensed images 
that correspond to the bright component q1 is $1.64$ arcsec, and the mean of
$\epsilon$ is $1.4\,$mas. This suggests that the actual position
uncertainty is much smaller than the value we assumed. Using so called the ``constant shift'' 
filtering function $W_{CS}(k)$ \citep{takahashi-inoue2014}, 
the square-root of the second
moment of $\eta$, $\langle \eta^2 \rangle^{1/2}=0.135 $ for 4 images,
Aq1, Bq1, Cq1, Dq1 has been obtained. The corresponding 
probability distribution function $p(\eta)$ with the theoretically 
obtained second moment can
be calculated using a fitting formula \citep{takahashi-inoue2014} 
that has been obtained from $N$-body and ray-tracing simulations
that can resolve haloes with a mass of $\sim 10^5\,\ms$ (for detail, see
section 3 in \citep{takahashi-inoue2014}).

\begin{figure}
\hspace{-0.18cm}
 \IG[width=85mm]{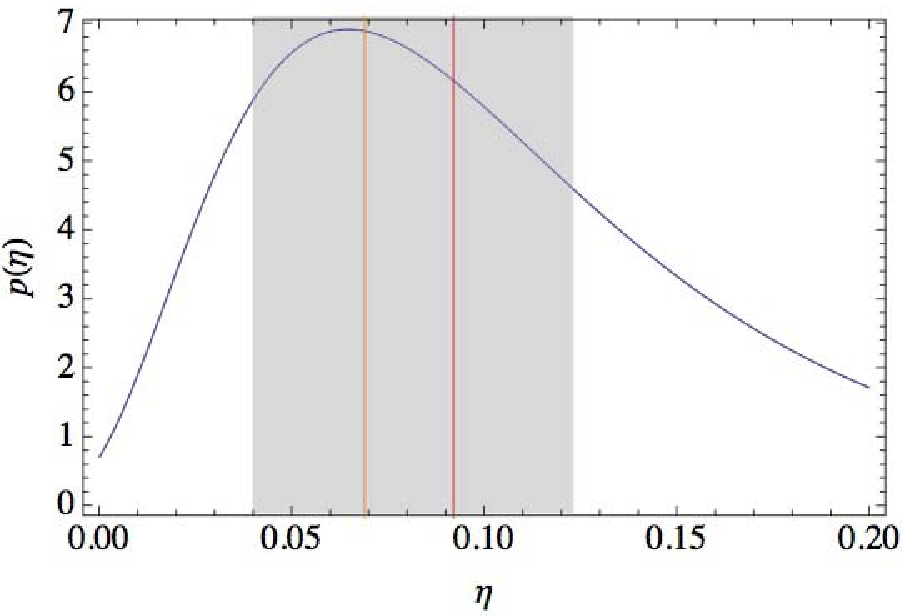}
\caption{Theoretical and observed magnification perturbation
 $\eta$. The blue curve corresponds to the probability distribution function of
 $\eta$ and the red and orange vertical lines correspond to the
estimators $\hat{\eta}$ for the best-fitted and concordant lens models
with aperture equal to $0.05$ arcsec. 
The grey region shows $1\,\sigma$ uncertainties due to observational
noises in the both lens models. 
 }
\label{PDF-of-eta}
\end{figure}

As shown in
Fig. \ref{PDF-of-eta}, $p(\eta)$ takes its peak at $\eta=0.065$.  One can see that the observed $\hat{\eta}$ 
shown in vertical lines (red and orange) with $1\,\sigma$ observational 
uncertainty (grey region) agrees with the prediction. The most
probable values are close to the peak. Thus the
observed anomalous aperture flux ratios in SDP.81 
can be explained by the line-of-sight structures.

\subsection{Homogeneous Spherical Clump}
As a simple model of a perturber in the line of sight, 
we consider a homogeneous spherical dark clump
that is compensated in mass. We assume that 
it has a positive uniform density $\rho_+$ inside 
radius $R_+$ and a negative uniform density $\rho_-$ at
$R_+<R<R_-$, where $R$ denotes the proper 
distance from the centre of a clump. At $R>R_-$, the density
is vanishing, i.e., $\rho=0$. The total mass at $R<R_+$ is
$M=4 \pi \rho_+ R_+^3/3$. Then, the $X$-component of 
the deflection angle $\hat{\alpha}_c$ at 
$\bvec{R}_\perp=(X,Y=0)$ in the lens plane 
is \citep{amendola1999},
\BE
\hat{\alpha}_c= b_c\times \left\{ 
\begin{array}{ll}
0, & \mbox{$ R_- \le |\bvec{R}_\perp|$} \\
\\
\tilde{X}^{-1}(\tilde{d}^3+3\tilde{d}^2+3\tilde{d})^{-1}
\\
\times \bigl[(1+\tilde{d})^2-\tilde{X}^2)\bigr]^{3/2}
, & \mbox{$R_+<|\bvec{R}_\perp|< R_-$} \\
\\
\tilde{X}^{-1}(\tilde{d}^3+3\tilde{d}^2+3\tilde{d})^{-1}
\\
\times 
\Bigl[ \bigl[(1+\tilde{d})^2-\tilde{X}^2)\bigr]^{3/2}
\\
-(1+\tilde{d})^3(1-\tilde{X}^2)^{3/2} \Bigr]
, & \mbox{$0\le |\bvec{R}_\perp| < R_+$},
\end{array}
\right.
\label{eq:void}
\EE
where $\tilde{X}\equiv X/R_+$, $\tilde{d}\equiv R_-/R_+-1$, 
and $b_c$ describes the mass scale, which is given by
\BEA
~~~~~~~~~~~~~~~~~~~~~b_c &\equiv& \frac{4 G M }{c^2 R_+}
\nonumber
\\
&=& \frac{16 \pi G R_c^2 \rho_+}{3 c^2}.
\EEA
The $Y$-component of $\hat{\bvec{\alpha}}_c$ is zero for $Y=0$.
For $Y\ne 0$, a rotation of coordinates by an angle 
$\phi=\tan^{-1}(Y/X)$ gives the deflection angle 
$\hat{\bvec{\alpha}}_c=(\hat{\alpha}_c,0)$
as the clump is spherically symmetric. 
The reduced deflection angle $\hat{\alpha}_c$ is 
a function of $\tilde{X}$.

Thus, the parameters of a spherically symmetric 
compensated homogeneous clump is the mass scale $b_c$, the width
of the shell $\tilde{d}$ and the proper radius of the positive 
density region $R_+$. $b_c$ represents the maximum possible 
astrometric shift. The reduced deflection angle $\hat{\bvec{\alpha}}_c$
vanishes at the centre of a clump and takes its peak at $R=R_+$.
For convenience, we choose the angular 
radius $\theta_+$ of the positive density region as the parameter
instead of $R_+$. In terms of the angular diameter radius $D$
of the clump, the angular radius is given by $R_+=D \theta_+$.
The position of the centre of a clump
in the image plane is denoted by $(x_c,y_c)$. 

\begin{figure}
\hspace{-0.1cm}
\IG[width=85mm]{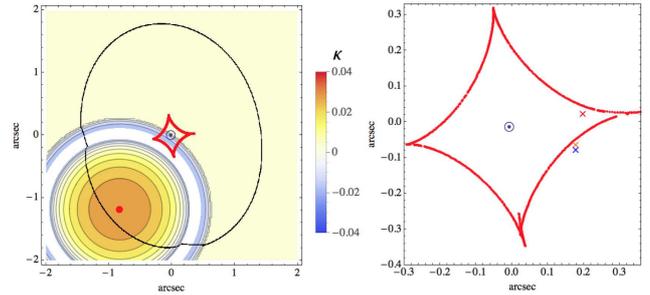}
\caption{Convergence $\kappa $ of a clump and 
critical curve (black) and caustic (red) for the best-fitted
 SIEES plus a clump model. The centre of the coordinates is at image
 G. In the left-hand panel, the red disk represents the centre of a compensated spherical clump
 in the line of sight. In the right-hand panel, the circled dot
 indicates the position of an SIE and crosses represent the positions of
 bright clumps in the source in the band 7 continuum image.}
\label{clump-caustic}
 \end{figure}

We study a toy model that consists of the best-fitted SIEES plus a homogeneous
compensated spherical dark clump. In order to explain the observed anomalous
flux ratios in the band 7 data and possible
astrometric shifts in the CO(8-7) data, we put a spherical clump centred
at image Cq1 in the band 7 data in the lens plane. 
After some trial and errors, we found a set of parameters that
can explain the both anomalous features (Table 5). The
convergences of the clump at the positive and negative density regions
are $\kappa = 0.014\pm 0.014$ and $\kappa = -0.015 \pm 0.0075$, 
respectively (see Fig. \ref{clump-caustic}). The mass inside $R_+=3.47
\textrm{kpc} h^{-1}$ is
$2.66 \times 10^9 \ms h^{-1}$ and the mean surface mass density inside $R_+$
is $6.76\times 10^8 \ms h^{-1} \tr{arcsec}^{-2}$. Because the gravitational potential of a
compensated clump vanishes at $R>R_-$, the effect on image A and image D is significantly small.

We also checked the expected amplitude of convergence perturbation due
 to line-of-sight dark structures using the semi-analytic method
 formulated in the Fourier space \citep{inoue-takahashi2012}. Assuming
 that the maximum wavelength of the perturbation  
is given by four times the radius of a clump $\theta_+$, we found that
the standard deviation is $\sqrt{\langle \kappa^2 \rangle}=0.020$. Thus, the expected
 amplitude agrees with the convergence perturbation due to the clump we
have considered.

In a similar manner described in section 4.2, we measured the inverted
aperture flux ratios in the band 7 based on the best-fitted or
concordant model with the clump.
As shown in Table \ref{clump-band7-flux}, 
the aperture flux ratios are consistent with being equal to 1
at the $0.5\,\sigma -1.5\,\sigma$ level. Comparing with the result in 
Table \ref{table-band7-flux}, one can see that the positive density
region within the clump induces a
decrease in the inverted flux of image C in the band 7, which has a negative parity and
the surrounding negative density shell also induces a decrease in that
of image B in the band 7, which has a positive parity. Such a decrease
cannot be expected for objects with a positive density provided that the
unperturbed model has a smooth potential and the first spatial
derivatives of the convergence, shear and magnification 
at the perturbed point are sufficiently
small\citep{inoue-takahashi2012, takahashi-inoue2014}.  
If one considers the effect of astrometric shifts, demagnification may be 
explained by shifting the image B towards the opposite direction to the nearest  
cusp in the source plane. However, if one imposes an accuracy of
$\sim 0.001$ arcsec in the position of q1, we find that the flux change would be
just a few percent, which is not sufficient for explaining the systematic
decrease of $\sim 20$ percent.

\begin{figure*}
\hspace{-0.00cm}
 \IG[width=170mm]{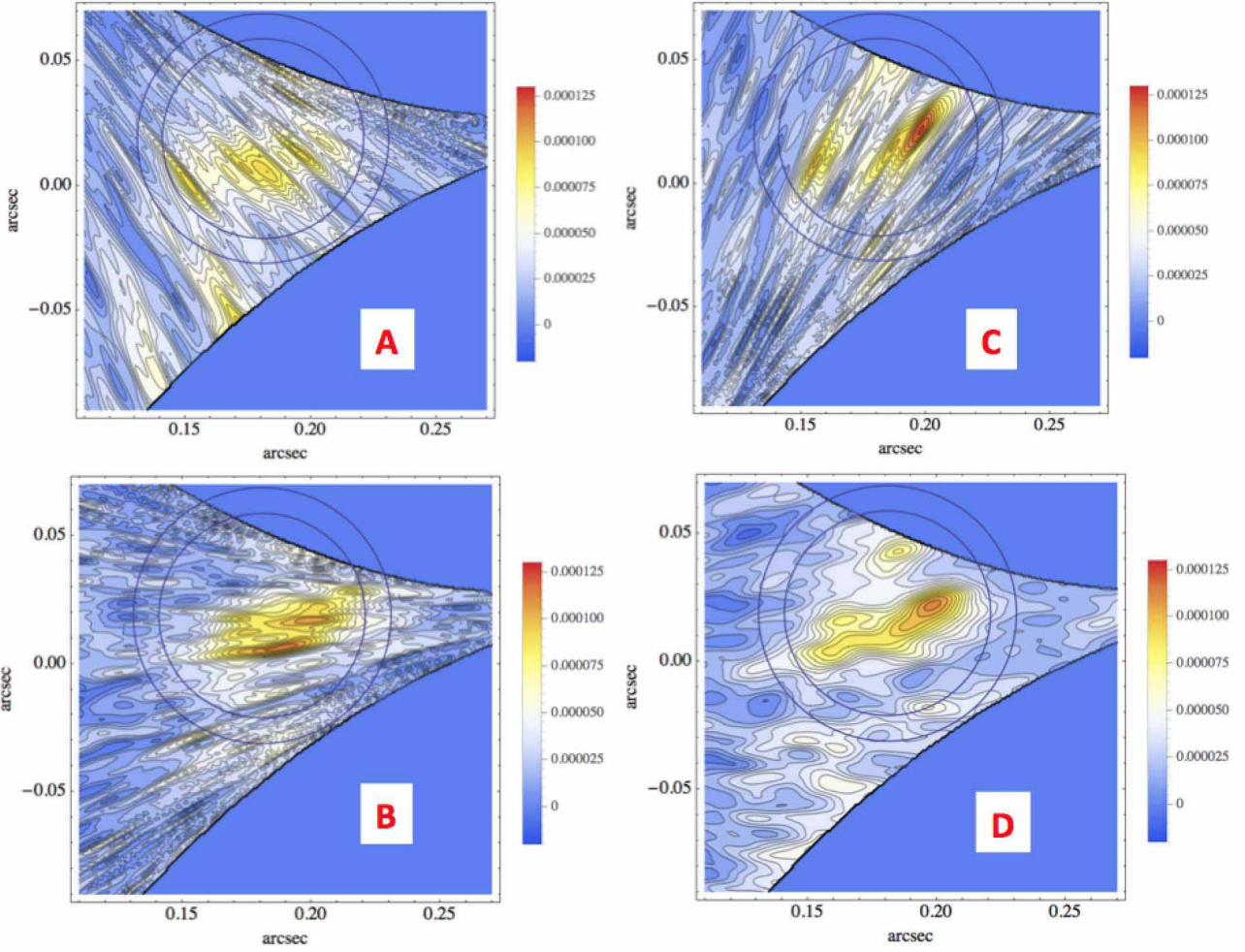}
\caption{Inverted quadruple images of band 7 data using a best-fitted SIEES
 plus a homogeneous compensated clump model. The unit is Jy per beam. 
Blue circles denote apertures with radii $0.04$ and $0.05$ arcsec. The centre of the apertures is 
$(0.1809, 0.01882)$ in the source plane.}
\label{clump-SIEES-inv}
\end{figure*}

\begin{figure*}
\hspace{-0.3cm}
 \IG[width=165mm]{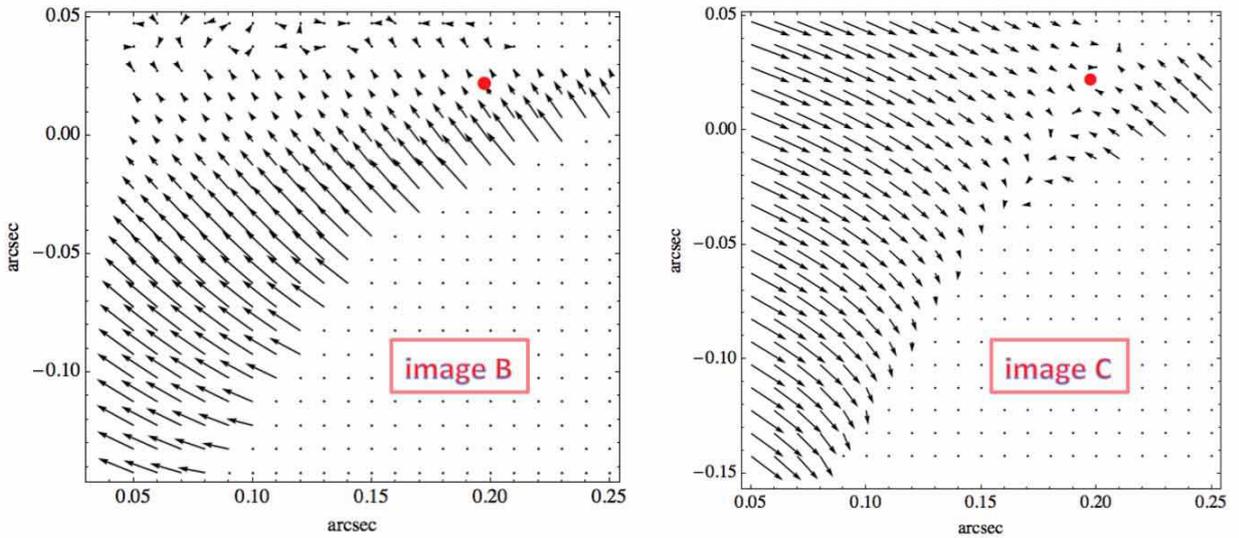}
\caption{Astrometric shifts in the source plane for image B and C. The arrows represent
the displacement of points due to a spherical compensated clump. Red discs
 (corresponding to the red disc in Fig. \ref{clump-caustic}) show the position of 
the centre of the clump. }
\label{shifts}
\end{figure*}

In Fig. \ref{clump-SIEES-inv}, we plot 
the surface brightness profiles of the inverted quadruple images of the band
7 data in the perturbed model. The changes in the aperture fluxes in
images B and C can be also explained by the fields of astrometric shift
of the order of 0.01 arcsec due to a clump (Fig. \ref{shifts}).

\begin{figure*}
\hspace{-0.46cm}
 \IG[width=165mm]{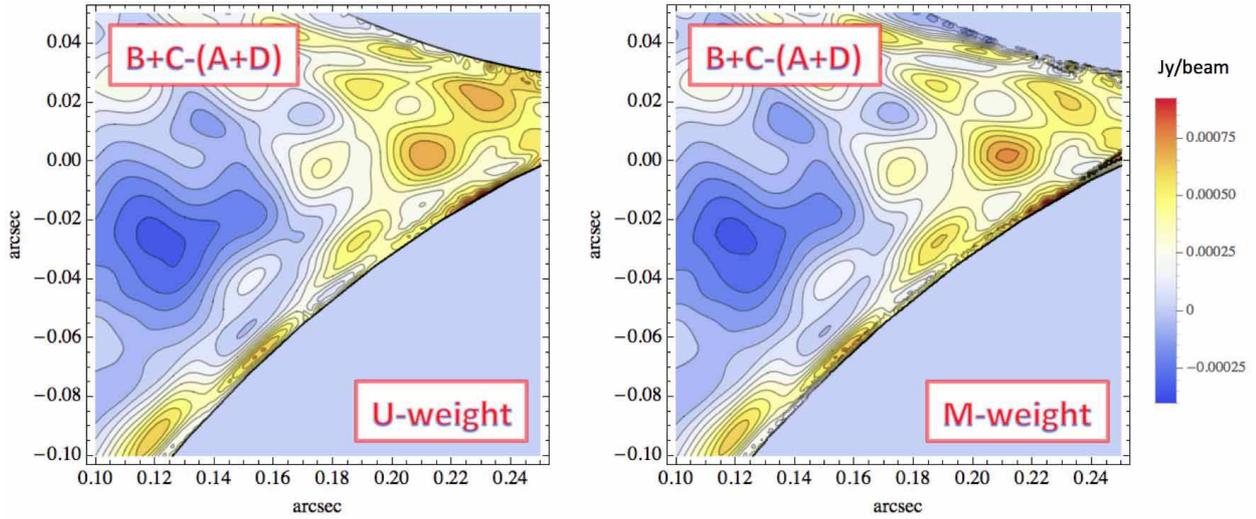}
\caption{Difference between the source images reconstructed from B+C
and A+D in the best-fitted model with a clump.  }
\label{clump-CO-diff}
\end{figure*}

\begin{figure*}
\hspace{-0.2cm}
 \IG[width=170mm]{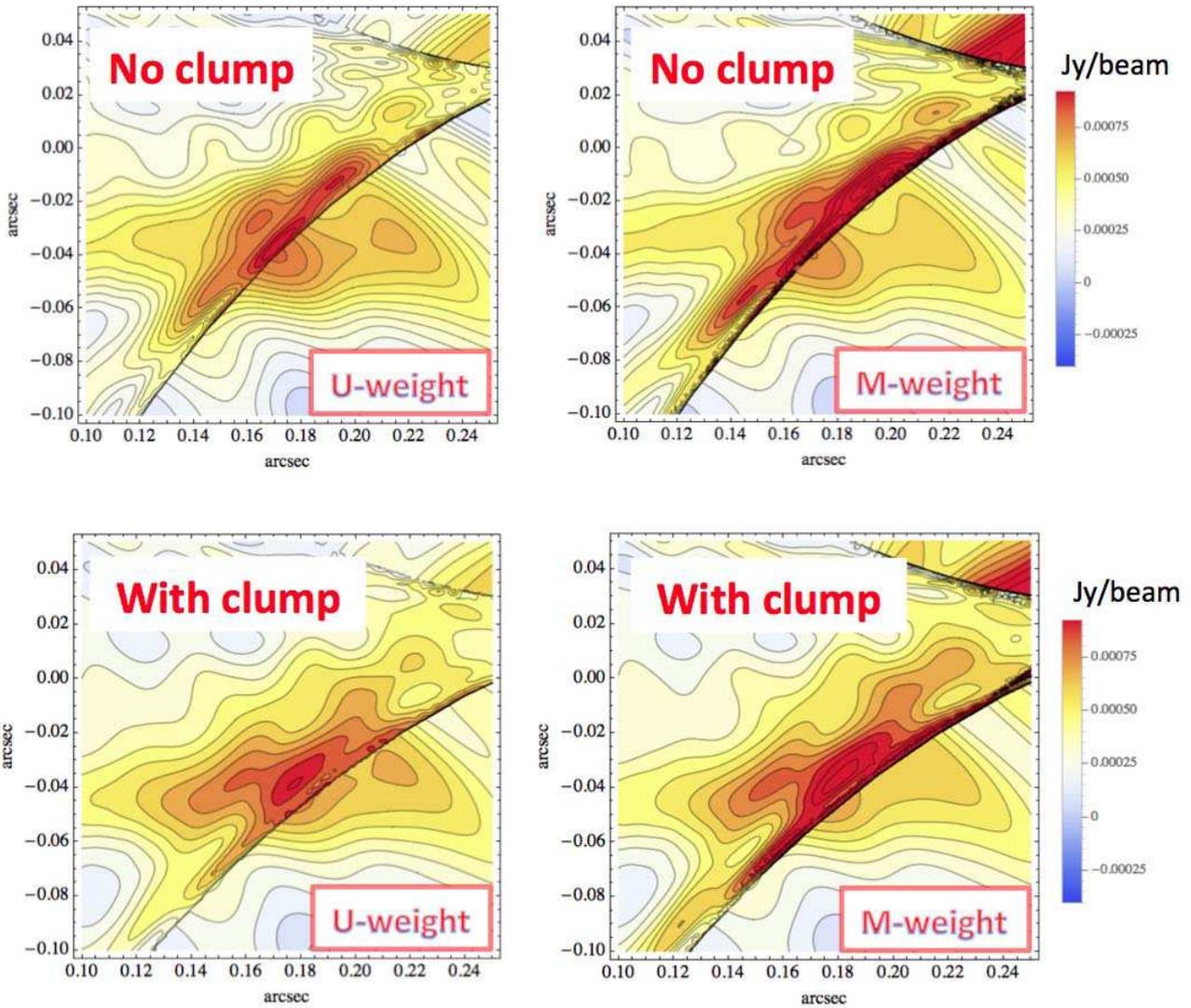}
\caption{Reconstructed source images of the CO(8-7) Ch.50 map 
 using the unperturbed (upper) and perturbed models (lower). }
\label{clump-CO-source}
\end{figure*}

The difference between
the inverted image reconstructed from B and C images (B+C) and 
that reconstructed from A and D images (A+D) in the CO(8-7) 50th channel data has
become small (Fig. \ref{clump-CO-diff}) due to astrometric
shifts of the order of 0.01 arcsec in the direction perpendicular to the
caustic (Fig. \ref{shifts}). As one can see in
Fig. \ref{clump-CO-source}, the reconstructed source images in
the CO(8-7) data in the neighbourhood of the caustic has become much smoother than
those in the model without a clump.
The observed asymmetric feature along with the caustic centred at a
bright clump has almost disappeared. Note that the remaining 
discontinuity at the edge of the caustic is due to the fact that the
beam size ($\sim 0.1$ arcsec) is much larger than the source size ($\sim
0.05$ arcsec). Even in the cases where the lens model is perfectly
correct, such a discontinuity is unavoidable.

\begin{figure*}
\hspace{-0.16cm}
 \IG[width=170mm]{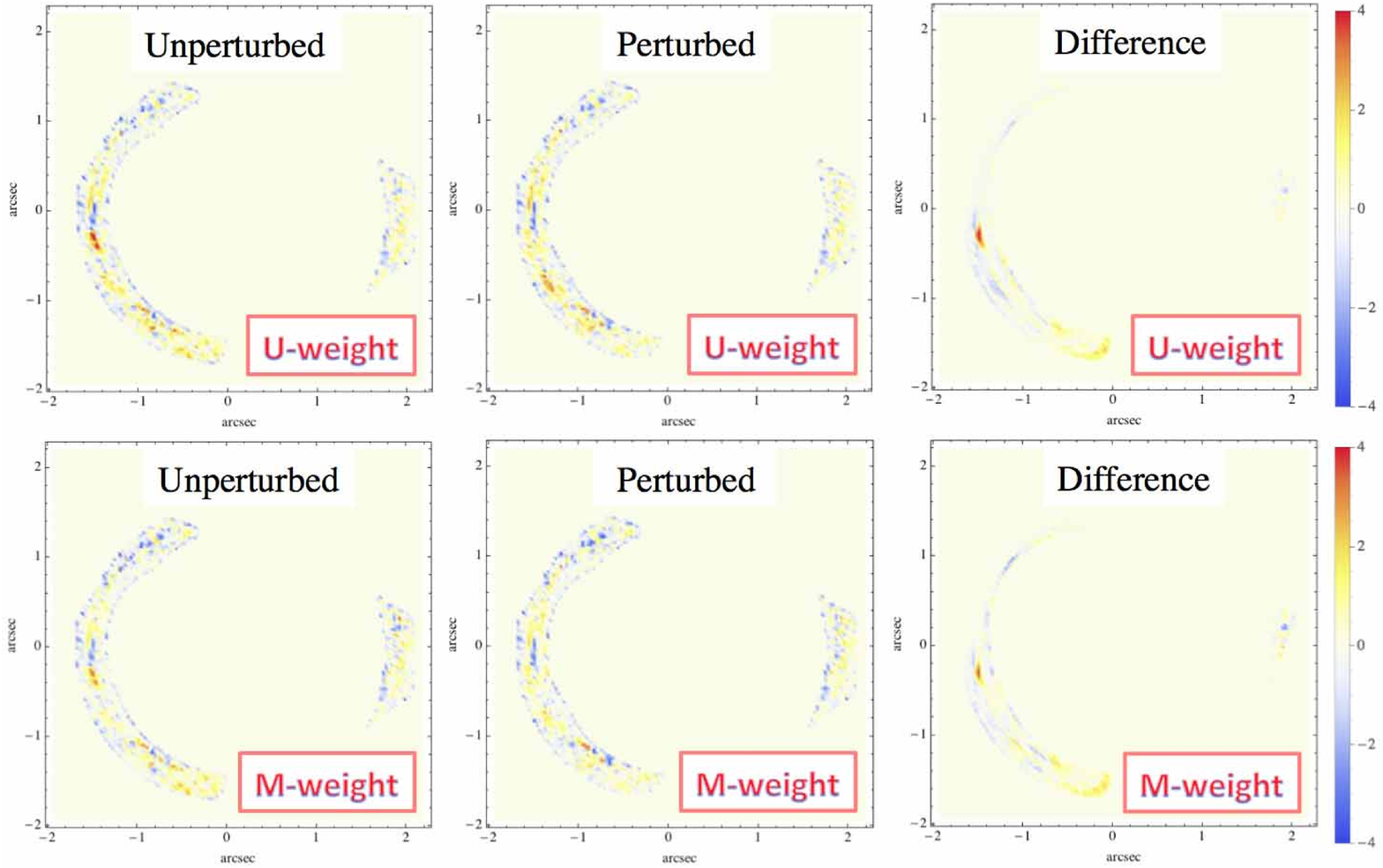}
\caption{The surface brightness 
differences with the observed band 7 image in the image plane 
(the model minus data). The pixel size is $0.03$ arcsec. The panels show 
the unperturbed (left) and perturbed (middle) best-fitted models and
their differences (right). The unit is $1\,\sigma$ in one pixel. }

\label{chi}
\end{figure*}

In Fig. \ref{chi}, we plot the difference between the model and
observed images in the band 7 continuum data for each pixel in the image plane.  
To account for the beam smoothing effect, the pixel size is downgraded to 
0.03 arcsec from 0.005 arcsec. We can see a prominent
difference that consists of two bright clumps 
(in red colour) in the neighbourhood of Bq1 at $(x,y)\approx
(-1.7,-0.3)$. The feature has become inconspicuous in the model with a
clump. Moreover, the difference in the neighbourhood of Cq1 centred at
$(x,y)\sim (-0.4,-1.5)$ has also become small in the model with a
clump. These modifications are consistent with those observed in the source
plane: the best-fitted model predicts too bright B and C images. In order
to explain the demagnification of B image, we need to consider a
perturbation by a negative density. 
\begin{table*}
\begin{center}
\caption{Parameters of a Spherical Homogeneous Clump}
  \begin{tabular}{c c c c c} 
\hline
  $(x_c,y_c)$ & $b_c$ (arcsec) & $\tilde{d}$ & $\theta_+$ (arcsec)  &
 $M(10^9 \ms/h)$ \\
\hline
 (-0.826,-1.184)  & 0.03 & 0.286  & 1.12 & 2.66  \\ 
\hline
  \end{tabular}
  \end{center}
\label{clump}
\end{table*}

\begin{table*}
\begin{center}
\caption{Modification of Flux Ratios in Band 7 }
  \begin{tabular}{c c c c c c c} 
\hline
  & aperture(arcsec) & $\eta(1\sigma)$ & & B/A & C/A & D/A \\
\hline
  &  &  & ratio  & 0.921 & 0.891  & 0.956  \\ \cline{4-7}
  &  0.06 &$0.043(0.039)$ &   error & 0.065 & 0.073 & 0.082  \\ \cline{4-7}
  &   & & deviation & 1.2 & 1.5 & 0.5   \\ \cline{2-7}
  &   &  & ratio  & 0.935 & 0.910  & 0.983  \\ \cline{4-7}
 best-fitted &  0.05 &$0.038(0.039)$ &  error & 0.062 & 0.073 & 0.078  \\ \cline{4-7}
 with a clump &   & & deviation & 1.1 & 1.2 & 0.2   \\ \cline{2-7}
  &  &  & ratio  & 0.938 & 0.926  & 0.996  \\ \cline{4-7}
  &  0.04 &$0.035(0.036)$ &   error & 0.059 & 0.071 & 0.074  \\ \cline{4-7}
  &   & & deviation & 1.1 & 1.0 & 0.0   \\ \hline
\\
\\
\hline
  & aperture(arcsec) & $\eta(1\sigma)$ & & B/A & C/A & D/A \\
\hline
  &  &  & ratio  & 0.937 & 0.919  & 0.980  \\ \cline{4-7}
  &  0.06 &$0.029(0.036)$ &   error & 0.071 & 0.067 & 0.078  \\ \cline{4-7}
  &   & & deviation & 0.9 & 1.2 & 0.25   \\ \cline{2-7}
  &   &  & ratio  & 0.957 & 0.934  & 1.000  \\ \cline{4-7}
 concordant &  0.05 &$0.030(0.037)$ &  error & 0.067 & 0.066 & 0.076  \\ \cline{4-7}
 with a clump &   & & deviation & 0.6 & 1.0 & 0.22   \\ \cline{2-7}
  &  &  & ratio  & 0.966 & 0.945  & 1.011  \\ \cline{4-7}
  &  0.04 &$0.034(0.036)$ &   error & 0.063 & 0.065 & 0.072  \\ \cline{4-7}
  &   & & deviation & 0.5 & 0.8 & 0.1   \\ \hline

\end{tabular}
\end{center}
\label{clump-band7-flux}
\end{table*}

\section{Conclusion and Discussion}
In this paper, we have analysed the ALMA long baseline science verification
 data of the gravitational lens system SDP.81. 
We have fitted the positions of the brightest clumps in the source and a possible AGN component of
 the lensing galaxy in band 7 continuum image using a canonical lens
 model SIEES. Then, we have measured
 the ratio of fluxes in some apertures defined 
at the source plane where the lensed
 images are inversely mapped. We have found 
that the aperture fluxes of B and C images in the band 7 
are demagnified by 10-20 percent in comparison with the A and D images 
with a significance at 2$\sigma$-3$\sigma$ level. We have observed an asymmetric
feature along the caustic in the CO (8-7) line at the 50th channel
 (rest-frame velocity 28.6\,$\textrm{km}\textrm{s}^{-1}$)
possibly due to astrometric shifts of the order of $0.01$ arcsec 
by some perturbers. Based on a semi-analytic calculation, we find that the
 magnitudes of observed anomalous flux ratios and the astrometric shifts can be
explained by dark structures on subgalactic scales in the line of sight. We have found that 
a spherical clump compensating in mass in the lens plane can explain the 
anomalous aperture flux ratios and an imprint of astrometric shifts. 
The redshift of the clump may be different from that of the primary
lens. It can be directly measured if
the astrometric shifts are measured with a good accuracy
 \citep{inoue-chiba2005a}. Note that the perturbers may 
have a more complex structure that consists of haloes, filaments and
 voids \citep{inoue2015}. 

One might be tempted to attribute the observed anomalous flux ratios to
 subhaloes or substructures in the lensing galaxy. However, we have found that a bright
 clump in image B with a positive parity isless magnified than
 the theoretical prediction in the band 7 continuum image. This feature cannot be
explained by a presence of a positive density perturbation as it causes 
magnification of an unperturbed image. By subtracting a constant
 convergence from the fitted model and adding a constant shear, one can obtain negative
surface density regions as well as positive ones without affecting 
the fluxes and positions of the fitted model due to the mass-sheet
 degeneracy. However, the amplitudes of convergence in such negative density regions are
 expected to be much smaller than those in positive density regions as 
 the outskirt of infalling subhaloes are stripped and the spatial
 correlation between the subhaloes is reduced due to tidal force in
 the host halo.

Instead, one can consider a possibility that 
the anomaly is caused by the line-of-sight
intergalactic structures. As is well known, the line-of-sight structures
 may consist of complex non-linear objects such as haloes, filaments, walls
 and voids. Therefore, the convergence 
perturbations consist of positive and negative density components 
which have a spatial correlation on scales of $\lesssim 10 \, \textrm{kpc}$.  
 Although the amplitude of negative components is somewhat smaller
 than that of positive components, the probability of 
 crossing with the photon path would be larger for negative 
components than positive ones. In order to find such perturbations, 
we need observations with a very good sensitivity and high resolution. ALMA
is an ideal tool for carrying out such observations.

The canonical lens model that we have considered may have been too simple.
Indeed, one can consider possibilities that the observed anomalies are due to
simplification of the unperturbed model. For instance, inclusion of
a core, a deviation from $\alpha=-2$ or a power law, higher multipoles, 
contribution from group galaxies or clusters may explain the anomalies. 
Indeed, increasing the number of model parameters would surely weaken the 
argument we have made. However, the observed 
anomalous feature in image B in the band 7 is localized in the
neighbourhood of the brightest clumps. This suggests that the fluctuation
scale of a possible perturbation is sufficiently smaller than the
effective Einstein radius of
the primary lens. Therefore, it is unlikely that modification of the
potential of the unperturbed model on large scales leads to the 
local change of magnification. Changing the parameters of the
best-fitted unperturbed model would not work unless one considers
a non-smooth potential that is not physically motivated.
In addition to the line-of-sight
structures, subhaloes in the lens galaxy may also perturb the 
flux ratios to some extent. In order to probe which contribution is dominant over
the other, observation of 21 cm absorption lines would be necessary
as intergalactic structures may retain a large amount of neutral
hydrogen gas.
 
We have not obtained an optimal solution for the 
perturbation of convergence and shear. The success of a simple 
toy model suggests that the perturbation consists of 
the both positive and negative density perturbations.
The actual shape of these perturbation might be obtained 
by more sophisticated methods formulated either in the $uv$ plane, image 
plane or source plane. It is also important to use
all the available data at different frequencies. More 
realistic ray-tracing simulations beyond the Born approximation 
are necessary for understanding the nature of flux anomalies and 
astrometric shifts due to intergalactic structures. These issues should 
be addressed in our future work. 

\section{Acknowledgements}
We thank Yoichi Tamura and Bunyo Hatsukade for useful discussions and 
comments. This work is supported in part by JSPS Grant-in-Aid for
Scientific Research (B) (No. 25287062) ``Probing the origin
of primordial minihaloes via gravitational lensing phenomena'',
MEXT Grant-in-Aid for Scientific Research on Innovative Areas
``Cosmic Acceleration''(No. 15H05889 for MC), and Ministry of 
Science and Technology (MoST) of Taiwan, MoST 103-2112-M-001-032-MY3.
This paper makes use of the following 
ALMA data: ADS/JAO.ALMA$\#$2011.0.00016.SV. ALMA is a partnership
of ESO (representing its member states), NSF (USA) and NINS
(Japan), together with NRC (Canada) and NSC and ASIAA (Taiwan), in
cooperation with the Republic of Chile. The Joint ALMA Observatory is
operated by ESO, AUI/NRAO and NAOJ.
\bibliographystyle{mnras}
\bibliography{weak-lensing-by-los2016}
%\bibliography{test}

\label{lastpage}
\end{document}